\documentclass{aa}  

\usepackage{txfonts}
\usepackage{textcomp}

\usepackage{graphicx}	
\usepackage{amsmath}	
\usepackage{amssymb}	

\usepackage[caption=false]{subfig}
\usepackage{booktabs}

\usepackage[normalem]{ulem}

\usepackage{xcolor}

\usepackage{footmisc}

\usepackage{natbib}
\usepackage{cleveref}
\usepackage{hyperref}

\usepackage{blindtext} 

\graphicspath{
	{plots/}
}

\begin{document}

    \title{The GIST Pipeline: A Multi-Purpose Tool for the Analysis and Visualisation of (Integral-field) Spectroscopic Data}
    \titlerunning{The GIST Pipeline}

    \author{%
        A. Bittner, 
        \inst{1,2}
        J. Falc\'on-Barroso, 
        \inst{3,4}
        B. Nedelchev, 
        \inst{5,6}
        A. Dorta, 
        \inst{3,4}
        D. A. Gadotti, 
        \inst{1}
        M. Sarzi, 
        \inst{5}
        A. Molaeinezhad,
        \inst{3,4}
        E. Iodice, 
        \inst{7}
        D. Rosado-Belza, 
        \inst{3,4}
        A. de Lorenzo-C\'aceres, 
        \inst{3,4}
        F. Fragkoudi,
        \inst{8}
        P. M. Galán-de Anta,
        \inst{9,5}
        B. Husemann, 
        \inst{10}
        J. M\'endez-Abreu, 
        \inst{3,4}
        J. Neumann,
        \inst{11}
        F. Pinna, 
        \inst{3,4}
        M. Querejeta,
        \inst{1,12}
        P. S\'anchez-Bl\'azquez
        \inst{13}\and
        M. K. Seidel
        \inst{14}
    }
    \authorrunning{A. Bittner et al.}

    \institute{%
        European Southern Observatory, Karl-Schwarzschild-Str. 2, D-85748 Garching bei M\"unchen, Germany\\ \email{adrian.bittner@eso.org}\and
        Ludwig-Maximilians-Universit\"at, Professor-Huber-Platz 2, 80539 Munich, Germany \and
        Instituto de Astrof\'isica de Canarias, Calle V\'ia L\'actea s/n, E-38205 La Laguna, Tenerife, Spain \and
        Departamento de Astrof\'isica, Universidad de La Laguna, E-38200 La Laguna, Tenerife, Spain \and
        Armagh Observatory and Planetarium, College Hill, Armagh, BT61 9DG, Northern Ireland, UK \and
        Center for Astrophysics Research, University of Hertfordshire, College Lane, AL10 9AB Hatfield, UK \and
        INAF-Osservatorio Astronomico di Capodimonte, via Moiariello 16, I-80131 Napoli, Italy \and
        Max-Planck-Institut f\"ur Astrophysik, Karl-Schwarzschild-Str. 1, D-85748 Garching bei M\"unchen, Germany \and
        Astrophysics Research Center, School of Mathematics and Physics, Queen's University Belfast, Belfast BT7 INN, UK \and
        Max-Planck-Institut fur Astronomie, Konigstuhl 17, D-69117 Heidelberg, Germany \and
        Leibniz-Institut f\"ur Astrophysik Potsdam (AIP), An der Sternwarte 16, D-14480 Potsdam, Germany \and
        Observatorio Astron\'omico Nacional (IGN), C/ Alfonso XII 3, E-28014 Madrid, Spain \and
        Departamento de F\'isica Te\'orica, Universidad Aut\'onoma de Madrid, E-28049 Cantoblanco, Spain \and
        Carnegie Observatories, 813 Santa Barbara St., CA 91101 Pasadena, USA
    }

    \date{Received XXXXX; accepted XXXXX}

    \abstract{%
    We present a convenient, all-in-one framework for the scientific analysis of fully reduced, (integral-field)
    spectroscopic data. The \texttt{GIST} pipeline (\textbf{G}alaxy \textbf{I}FU \textbf{S}pectroscopy \textbf{T}ool) is
    entirely written in Python~3 and conducts all steps from the preparation of input data, over the scientific analysis
    to the production of publication-quality plots. In its basic setup, it extracts stellar kinematics, performs an
    emission-line analysis and derives stellar population properties from full spectral fitting as well as via the
    measurement of absorption line-strength indices by exploiting the well-known \texttt{pPXF} and \texttt{GandALF}
    routines, where the latter has now been implemented in Python. The pipeline is not specific to any instrument or
    analysis technique and provides easy means of modification and further development, as of its modular code
    architecture.  An elaborate, Python-native parallelisation is implemented and tested on various machines.  The
    software further features a dedicated visualization routine with a sophisticated graphical user interface.  This
    allows an easy, fully-interactive plotting of all measurements, spectra, fits, and residuals, as well as star
    formation histories and the weight distribution of the models.
    The pipeline has successfully been applied to both low and high-redshift data from MUSE, PPAK (CALIFA), and SINFONI,
    as well as to simulated data for HARMONI@ELT and WEAVE and is currently being used by the TIMER, Fornax3D, and
    PHANGS collaborations. We demonstrate its capabilities by applying it to MUSE TIMER observations of NGC\,1433. 
    }


    \maketitle


\section{Introduction}
\label{sec:introduction}
Over the past decades, spectroscopic observations have provided significant insights into the fundamental properties of
galaxies. In particular, the measurement of stellar and gaseous motions as well as the inference of stellar population
properties have made substantial contributions to our understanding of the formation and evolution of galaxies. 

With the introduction of the first integral-field spectrographs \citep[IFS, see e.g.][]{bacon1995}, it became feasible
to perform such observations in a spatially resolved manner and for larger samples of galaxies.  The SAURON survey
\citep{bacon2001,deZeeuw2002} was one of the first projects to make extensive use of this technology.  Based on their
representative sample of 72 nearby early-type galaxies, the project has investigated stellar and gaseous kinematics
\citep{emsellem2004,sarzi2006,jfb2006,ganda2006}, stellar population properties \citep{peletier2007,kuntschner2010}, and
eventually distinguished fast and slow rotating early-type galaxies \citep{emsellem2007}. 
Subsequently, the ATLAS$^{\mathrm{3D}}$ project \citep{cappellari2011} has continued this endeavour by further
investigating kinematic properties, for instance the global specific angular momentum \citep{emsellem2011}, based on a
volume complete sample of 260 early-type galaxies. 
%
The CALIFA survey \citep{sanchez2012} advanced these previous studies towards a morphologically unbiased sample of 667
galaxies. In addition, it provides a unique combination of a large spatial coverage of a few effective radii and high
spatial sampling \citep[see also][]{garcia-benito2015}. Other surveys aim at observing even larger samples: while the
SAMI survey \citep{croom2012,bryant2015} includes $\sim 3000$ galaxies across different environments, MaNGA
\citep{bundy2015} will contain approximately 10\,000 nearby galaxies. 
%
Most recently, these projects are being complemented by IFS studies of local galaxies in unprecedented spatial
resolution \citep[e.g. MUSE,][]{bacon2010}. For instance, the TIMER project \citep{gadotti2019} analyses the central
structures of 24 barred local disc galaxies, in order to study the formation histories of these structures and infer
constraints on the related secular evolution processes. Fornax3D \citep[][Iodice et al. submitted]{sarzi2018}
investigates mostly early-type galaxies in the Fornax cluster environment, spatially covering galaxies up to four
effective radii.  The PHANGS survey (Leroy et al.\footnote{\url{http://www.phangs.org/}}, in prep.) aims to connect the
physics of gas and star formation with the large-scale galactic structure by complementing IFS data from MUSE with
interferometric data from ALMA. 

Spatially resolved spectroscopic data contain an outstanding amount of information. It is not only possible to measure
the stellar motions and perform an emission-line analysis, but also to infer star-formation histories.  However, the
extraction of such quantities from the science-ready IFS data is very complex and sophisticated analysis techniques are
indispensable. 
Over time, a set of commonly used and well-tested techniques for this analysis has emerged.  After various pioneering
works on the measurement of stellar kinematics since the early 1970s \citep[see
e.g.][]{simkin1974,sargent1977,tonry1979,franx1989,bender1990,rix1992,kuijken1993,vanderMarel1993,statler1995,merritt1997},
the large majority of recent studies exploit the penalised pixel-fitting code \texttt{pPXF}
\citep{cappellari2004,cappellari2017} to infer a stellar line-of-sight velocity distribution (LOSVD). 
Similarly, the ``Gas and Absorption Line Fitting'' (\texttt{GandALF}) software \citep{sarzi2006,jfb2006} as well as the
\texttt{pyPARADISE} tool \citep{walcher2015,husemann2016} are widely-used methods to conduct a thorough emission-line
analysis. 
Non-parametric star-formation histories can, for instance, be extracted with \texttt{MOPED} \citep{moped},
\texttt{pPXF}, \texttt{STARLIGHT} \citep{starlight}, \texttt{STECKMAP} \citep{ocvirk2006b,ocvirk2006a}, \texttt{VESPA}
\citep{vespa}, as well as \texttt{ULySS} \citep{ulyss}.

Such a well-established set of sophisticated techniques provides a good base for the analysis of spectroscopic data.
Moreover, the application of these techniques to actual data is, in principle, not difficult, as implementations of
these codes are widely available.  However, in practice this task soon becomes very tedious: the use of different IFS
instruments, spectral template libraries, and analysis setups in the context of different scientific objectives and
surveys add some inconvenience to the day-to-day usage of these techniques. Further complications arise if a very large
sample of galaxies makes the use of an automated pipeline necessary, especially in light of the tremendous amount of
spectra provided by state-of-the-art instruments (e.g. MUSE) and therefore the inevitable requirement of parallelised
software. 

We aim to overcome these inconveniences by introducing the \texttt{GIST} pipeline (\textbf{G}alaxy \textbf{I}FU
\textbf{S}pectroscopy \textbf{T}ool) that is modular and general enough to be easily applied to data from all existing
IFS instruments as well as in the context of different scientific objectives and surveys. In particular, the code
architecture provides easy means of modification and expansion, while being a convenient all-in-one framework for the
scientific analysis of fully reduced IFS data. In addition, the software package features sophisticated visualization
routines and employs a well-tested parallelisation. While this is not the first time a pipeline of this sort has been
developed (e.g. \texttt{Pipe3D}, \citealt{sanchez2016}; \texttt{LZIFU}, \citealt{i-ting2016} or the MaNGA pipeline,
\citealt{westfall2019}), it is the first one offered publicly with a wide range of built-in configurations for several
instruments and surveys.
The large scientific community that is in need of such a multi-purpose IFS analysis tool is highlighted by the fact that
the \texttt{GIST} pipeline is already being employed by the TIMER, Fornax3D, and PHANGS collaborations.  In addition,
various researchers are applying the code to both low and high-redshift data from MUSE, PPAK (CALIFA), and SINFONI
\citep{sinfoni}, as well as to simulated data for HARMONI@ELT \citep{harmoni} and WEAVE IFU modes \citep{weave}. 

This paper is organised as follows: in the next section we introduce the core design principles of the pipeline, its
code architecture and workflow, as well as the most important aspects of the implementations. In
Sect.~\ref{sec:application} we apply the pipeline to TIMER data of NGC\,1433 obtained with the MUSE spectrograph.  We
finish with a summary of this work in Sect.~\ref{sec:conclusion}.

\section{The Pipeline}
\label{sec:code}
\subsection{Core Design Principles}
\label{subsec:coreDesignPrinciples}
The specific pipeline architecture is designed with the following high-level objectives: 
\begin{description}
    \item[\textbf{Convenience:}] 
        We aim to provide a convenient, all-in-one framework for the scientific analysis of fully reduced
        (integral-field) spectroscopic data. This includes all steps from the read-in and preparation of input data over
        the several analysis steps to the production of well-structured output tables and publication-quality plots.
        Such a framework not only accelerates and simplifies the analysis, but also assures the consistency of the
        analysis throughout large data samples. 
    \item[\textbf{Extensive Functionality:}]
        The basic version of the \texttt{GIST} pipeline extracts stellar kinematics and non-parametric star formation
        histories by exploiting the \texttt{pPXF} routine \citep{cappellari2004,cappellari2017}, determines gaseous
        kinematics and emission-line fluxes with the \texttt{GandALF} procedure \citep{sarzi2006,jfb2006}, and measures
        line strength indices as well as the corresponding single-stellar population equivalent population properties
        \citep[SSP;][]{kuntschner2006,martin-navarro2018}. These methods can be executed in a multitude of different
        flavours and configurations (see Sect.~\ref{subsubsec:configuration}). 
    \item[\textbf{Flexibility:}] 
        This code is not specific to any instrument, fitting procedure or scientific project.  Instead, the
        implementation features a high level of flexibility, as we aim to provide an analysis framework that suits the
        needs of a variety of scientific objectives across various collaborations.  This flexibility becomes mostly
        evident through the fact that the pipeline is capable of handling data from different instruments and its
        execution can be tailored to reflect a variety of different flavours in either interactive or non-interactive
        batch processing manner.
    \item[\textbf{Modifiability/Expandability:}] 
        Regardless of how extensive and flexible an analysis pipeline is, it is virtually impossible that one particular
        design satisfies all conceivable scientific requirements. Thus, it is necessary to provide easy means of
        modification of the code that will allow to tackle specific problems. This can be achieved by using a modular
        architecture while keeping the source code as clean and readable as possible. Moreover, this makes it
        straightforward for the user to expand the pipeline by adding user-defined modules. 
    \item[\textbf{Visualization:}] 
        The \texttt{GIST} pipeline outputs various high-level data products. However, while visualization routines for
        input IFS cubes are widely available (e.g.
        QFitsView\footnote{\url{http://www.mpe.mpg.de/~ott/dpuser/qfitsview.html}}), the amount of software to display
        the data products is still limited (e.g. E3D\footnote{\url{http://www.caha.es/sanchez/euro3d/}},
        PINGSoft\footnote{\url{https://www.inaoep.mx/~frosales/pings/html/software/}},
        MARVIN\footnote{\url{https://www.sdss.org/dr15/manga/marvin/}}), and it is often specific for an instrument or
        project.  Such visualization routines are important though to simplify the inspection of the output and
        facilitate the immediate monitoring of observed spectra, their best fits and residuals.
        Thus, in this software package we provide the dedicated visualization routine \texttt{Mapviewer}, specifically
        designed to access \emph{all} data products of this pipeline.  To this end, it exploits a sophisticated
        graphical-user interface with fully interactive plots. For instance, the spectra, its corresponding fits and
        residuals, as well as star formation histories and the weight distribution of the models in a particular bin can
        be plotted by only a simple mouse-click on the given bin in the map. The \texttt{Mapviewer} is highly
        performance optimised and thus capable of handling the bulk of data products without noticeable interruption.
    \item[\textbf{Performance:}] 
        State-of-the-art IFS instruments output an enormous number of spectra per pointing. This implies significant
        computational expenses for the analysis of the data, in particular when larger samples of galaxies are
        considered. This necessitates the use of parallelised software and large, cluster-scale computing resources. In
        particular, a Python-native parallelisation, based on its \texttt{multiprocessing} module, is implemented to
        conduct the analysis on multiple spectra simultaneously. This parallelisation has been successfully tested with
        up to 40 processors on various machines and scales linearly. 
\end{description}

The code is publicly available through a dedicated
webpage\footnote{\url{https://abittner.gitlab.io/thegistpipeline}\label{webpage}}.  Additional analysis modules, for
instance the inclusion of \texttt{STECKMAP} and \texttt{pyPARADISE}, will be done in the future. We further encourage
the community to put forward their own input on the future development of the code, in particular by creating their own
specific modules. 

This paper intends to give an overview of the extensive capabilities of the code and is complemented by online
documentation\textsuperscript{\ref{webpage}}, which includes detailed information on the practical use of the pipeline.
Therefore we describe its download, installation, as well as a dedicated tutorial. In addition, the documentation
provides various examples of the pipeline usage and instructions on its modification. 

Although we provide this pipeline as a convenient, all-in-one framework for the analysis of IFS data, it is of
fundamental importance that the user understands exactly how the involved analysis methods work.  \emph{We warn that the
improper use of any of these analysis methods, whether executed within the framework of the \texttt{GIST} pipeline or
not, will likely result in spurious or erroneous results and their proper use is solely the responsibility of the user.
Likewise, the user should be fully aware of the properties of the input data before intending to derive high-level data
products.}  Therefore, this pipeline should not be simply adopted as a ``black-box''. To this extend, we urge any user
to get familiar with both the input data and analysis methods, as well as their implementation in this pipeline.

\subsection{Pipeline Architecture and Workflow}
\label{subsec:workflow}
\begin{figure}
    \includegraphics[width=\hsize]{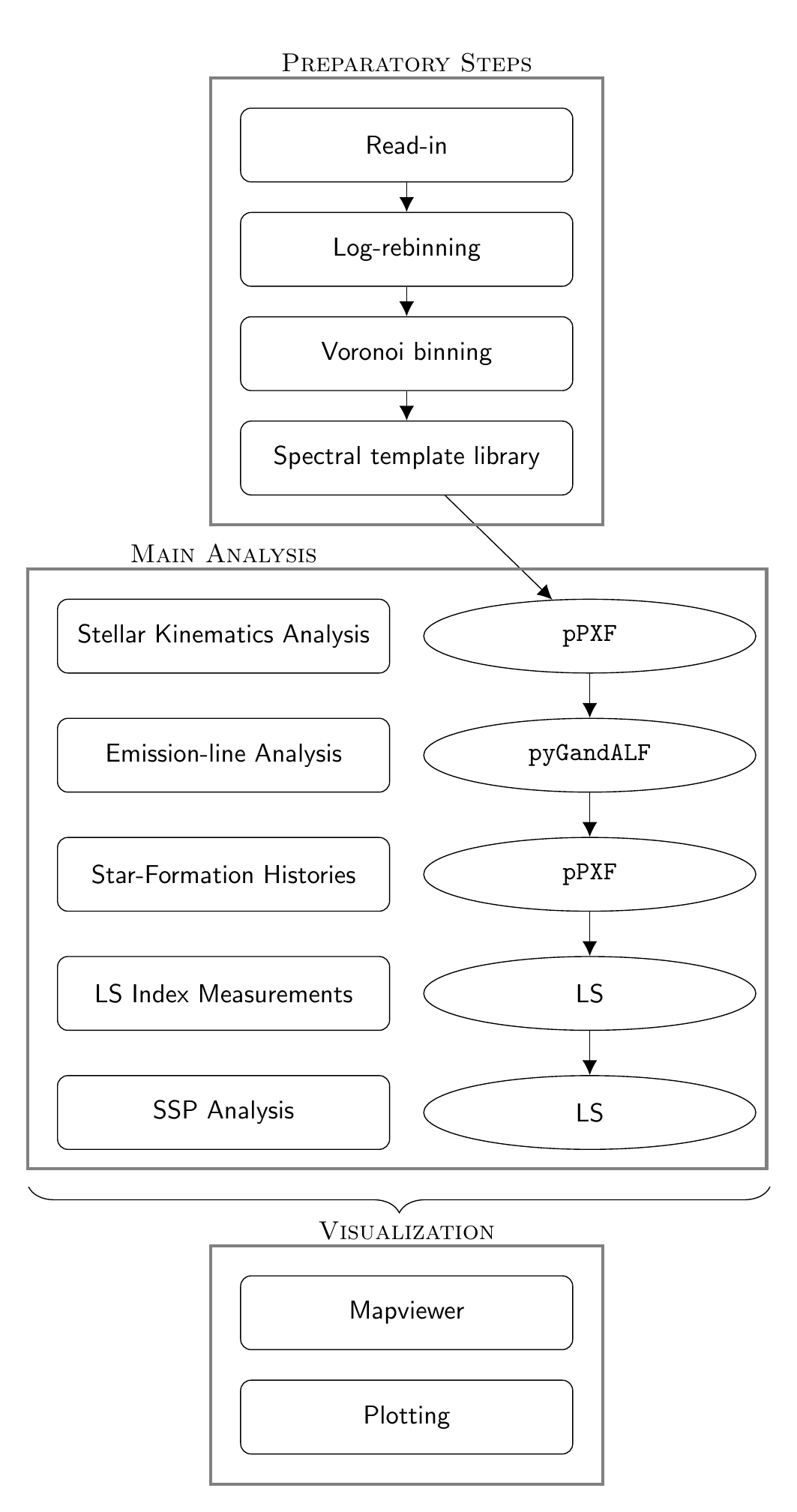}
    \caption{%
        Flowchart illustrating the architecture of the pipeline. The code is structured in three parts: preparatory
        operations, main analysis procedures, and visualization routines. Each of these part consists of various
        individual modules.
    }
    \label{fig:flowchart}
\end{figure}

To achieve the design goals presented in the previous section, the pipeline is structured in three major elements:
preparatory operations, main analysis procedures, and visualization routines. Each of these parts consists of various
individual modules. Figure~\ref{fig:flowchart} illustrates the code architecture. In
Sect.~\ref{subsec:pipelineImplementation} we present an in-depth discussion of the implementation of each of these
modules. 

The four major preparatory steps are the read-in of the fully-reduced, science-ready data cube, the logarithmic
rebinning of the spectra in the wavelength dimension (to obtain spectra that are linearly binned in velocity space), the
Voronoi binning of the spectra in the spatial dimensions (using the routine of \citealt{cappellari2003} to obtain
integrated spectra of approximately constant signal-to-noise ratio), as well as the necessary preparation steps in
regards to the spectral template library. There preparatory steps need be executed regardless of any further analysis
and are not repeated afterwards. 

The main analysis section consists of four modules: two of such modules exploit the \texttt{pPXF} routine in order to
extract stellar kinematics (hereafter ``PPXF-module'') and non-parametric star formation histories (denoted
``SFH-module''). Emission-line kinematics and fluxes are extracted by calling a new Python implementation of the
original \texttt{GandALF} procedure (denoted ``GANDALF-module'').  Finally, line strength indices and their
corresponding SSP-equivalent population properties are measured with the routines already used by \citet{kuntschner2006}
and \citet{martin-navarro2018} (denoted ``LS-module'').  These analysis modules are simply wrappers around the
implementations of the mentioned analysis techniques, mainly acting as an interface to the pipeline and handling the
input and output.  These four modules are fully independent, but may use the results of one another as input. In other
words, the modules can be turned on and off, or even be replaced by different, user-defined modules without affecting
the overall integrity of the pipeline.  As these modules are independent, the parallelisation is implemented in each
module individually (see Sect.~\ref{subsubsec:configuration} for a full discussion on the configuration options).

\begin{figure*}
    \includegraphics[width=\hsize,trim={1.2cm 1.2cm 1.2cm 1.2cm}]{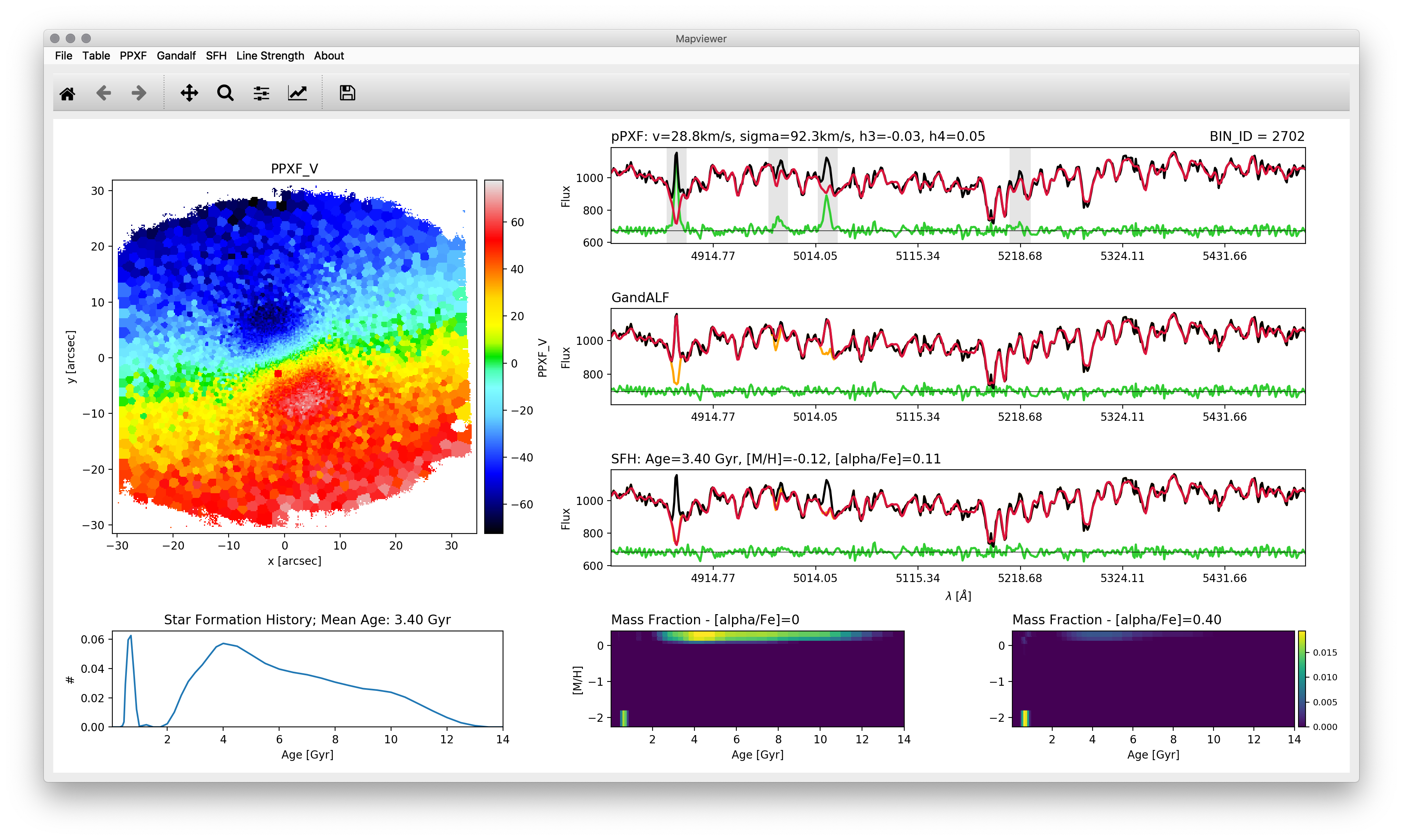}
    \caption{%
        Screenshot of the visualization software \texttt{Mapviewer} in its default layout, illustrating the results of
        the galaxy NGC\,1433. High-level data products of previous analysis runs of the \texttt{GIST} pipeline can be
        selected and displayed through the items in the menu bar. 
        Most importantly, a map of the selected quantity is displayed in the upper left panel of the window.  Individual
        bins in this map can be selected by simply performing a mouse-click on the map or providing the bin ID. The
        selected bin is subsequently marked by a blue circle while the other panels display more detailed information. 
        In particular, the upper three panels on the right-hand side of the window display observed spectra, and the
        fits generated by the PPXF-module (first panel), GANDALF-module (second panel) and SFH-module (third panel).
        Observed spectra are displayed in black, fits in red, emission-subtracted continuum spectra in orange, and
        residuals in green.  Masked spectral regions are highlighted in grey.  The three bottom panels display further
        information on the stellar population properties: non-parametric star formation histories are displayed in the
        left-hand panel while the centre and right-hand panels illustrate the weights of the templates in the
        age-metallicity-alpha grid. 
        The corresponding quantitative results of the measurements are provided next to each panel. 
        Different layouts of the \texttt{Mapviewer} routine are available to highlight the results of the line strength
        module.
    }
    \label{fig:mapviewerScreenshot}
\end{figure*}

Finally, visualization routines complement the analysis framework. Publication-quality plots of all results are
automatically generated during the analysis. In addition, all plotting routines can also be executed independently from
the analysis pipeline in order to allow the user any specific plotting preferences. A fully interactive visualization
method is provided by the routine \texttt{Mapviewer}, as discussed in Sect.~\ref{subsec:coreDesignPrinciples}.  A
screenshot of this routine is displayed in Fig.~\ref{fig:mapviewerScreenshot}. 

In addition, the pipeline workflow is optimised in two further ways: firstly, if the user has set unreasonable
configuration options, the pipeline will print a warning message and skip the affected module or, if unavoidable, the
analysis of the galaxy in consideration.  Secondly, the code checks if a part of the results is already available in the
output directory, because of, for instance, a previous partial run of the \texttt{GIST} pipeline. If such outputs are
detected, the corresponding analysis modules are not executed again. 

Such an automatic determination of whether or not a module has been already executed results in several advantageous
features: firstly, it allows to stop and later continue the workflow after each analysis module. For instance, it is
possible to first extract stellar kinematics for a given sample of galaxies, then the gaseous kinematics in a second
step, and finally any of the population properties without unnecessarily repeating any operation.  Furthermore, this
allows the validation of the outcome of any of the previous steps of the workflow before its continuation.  In other
words, this makes it possible to use the pipeline in a highly interactive manner which might be helpful for small
samples with a large number of spectra per galaxy. 

Secondly, this ensures the stability of the workflow against any unforeseen problems (i.e. numerical issues).  This is
of particular importance if one intends to use the code as a survey-level pipeline on a large sample of galaxies. The
survey-level usage of the pipeline is further supported by the fact that the configuration file can, in principle, be
machine generated and contain an arbitrarily large number of objects. 

Thirdly, it supports the reutilisation of intermediate results. If the user plans to conduct several test runs on the
same galaxy to e.g.\ quantify the impact of the adopted line-spread-function on the stellar kinematics, the results of
the preparatory operations can be re-used.  Similarly, these intermediate results can be modified to perform the
analysis in a different manner. For instance, one can modify the Voronoi map to conduct the analysis in manually defined
apertures instead of Voronoi bins.

\subsection{Pipeline Implementation}
\label{subsec:pipelineImplementation}
%
\subsubsection{Configuration}
\label{subsubsec:configuration}
To provide a maximum of convenience to the user, most of the configurations of the \texttt{GIST} pipeline are defined in
one file. This file contains the main switches that state which analysis modules need to be executed, as well as the
general settings and parameters, such as the used instrument, the wavelength range in consideration, the target
signal-to-noise ratio for the Voronoi binning, and the selection of the spectral template library. In addition, more
specific settings for any of the four analysis modules are provided. An overview about all configuration switches and a
brief description of their function is presented in Table~\ref{tab:configs}. An in-depth description of the effect of
these switches follows in this section. 

The line-spread-function (LSF) of the data is stated by the user in a configuration file that simply contains two
columns, defining the spectral resolution of the data at a given wavelength. The information is read-in and a linear
interpolation function is generated. Thus, it is possible to reconstruct the spectral resolution at any wavelength. The
spectral resolution of the template spectra is handled alike. Spectral regions which are masked in the analysis of the
PPXF and SFH modules, are defined by the user in distinct files stating the wavelength and width of the spectral masks,
respectively.

In addition to these files, there are only two more configuration files necessary: one specifying the emission-lines to
be masked or fitted during the analysis of the GANDALF-module and a file which states the wavelength bands of the
absorption line strength measurements.  These configurations will be explained in detail in the context of their
analysis modules below. 

\begin{table}
    \centering
    \begin{tabular}{p{2.15cm}p{6cm}}
    \toprule
    \toprule
    Config-Switch             &  Brief Description \\ 
    \midrule
    \multicolumn{2}{p{6cm}}{\emph{Main Switches}} \vspace{0.1cm} \\
    \texttt{DEBUG}            &  Run pipeline on one, central line of pixels \\
    \texttt{VORONOI}          &  Voronoi-bin the data prior to the analysis \\
    \texttt{GANDALF}          &  Run the GANDALF-module (\texttt{pyGandALF}) on bins or spaxels \\
    \texttt{SFH}              &  Run the SFH-module (regularised \texttt{pPXF})\\
    \texttt{LINE\_STRENGTH}   &  Run the LS-module\\
    \texttt{PARALLEL}         &  Activate multiprocessing \\
    \texttt{NCPU}             &  Number of cores to use for multiprocessing \\
    \midrule
    \multicolumn{2}{p{6cm}}{\emph{General Settings}} \vspace{0.1cm} \\
    \texttt{RUN\_NAME}        &  Name of the analysis run \\
    \texttt{IFU}              &  Identifier of the read-in routine \\
    \texttt{LMIN}             &  Minimum restframe wavelength [$\AA$]\\
    \texttt{LMAX}             &  Maximum restframe wavelength [$\AA$]\\
    \texttt{ORIGIN}           &  Origin of the coordinate system [arcsec] \\
    \texttt{REDSHIFT}         &  The redshift of the system [z] \\
    \texttt{SIGMA}            &  Initial guess of velocity dispersion [km/s]\\
    \texttt{TARGET\_SNR}      &  Target signal-to-noise ratio for the Voronoi binning\\
    \texttt{MIN\_SNR}         &  Minimum signal-to-noise ratio per spaxel to be accepted for the Voronoi binning\\
    \texttt{COVAR\_VOR}       &  Correct for spatial correlations of the noise in the Voronoi binning process\\
    \texttt{SSP\_LIB}         &  Defines the spectral template library\\
    \texttt{NORM\_TEMP}       &  Normalise the spectral template library to obtain light- or mass-weighted results\\
    \midrule
    \multicolumn{2}{p{6cm}}{\emph{PPXF Settings}} \vspace{0.1cm} \\
    \texttt{MOM}              &  Number of kinematic moments\\
    \texttt{ADEG}             &  Degree of the add. Legendre polynomial \\
    \texttt{MDEG}             &  Degree of the mult. Legendre polynomial \\
    \texttt{MC\_PPXF}         &  Number of Monte-Carlo simulations used to extract errors on the stellar kinematics\\
    \midrule
    \multicolumn{2}{p{6cm}}{\emph{GANDALF Settings}} \vspace{0.1cm} \\
    \texttt{FOR\_ERRORS}      &  Derive errors on the emission-line analysis \\
    \texttt{REDDENING}        &  Include the effect of reddening by dust \\
    \texttt{EBmV}             &  De-redden the spectra for the Galactic extinction in the direction of the target\\
    \midrule
    \multicolumn{2}{p{6cm}}{\emph{SFH Settings}} \vspace{0.1cm} \\
    \texttt{REGUL\_ERR}       &  Regularisation error (The reciprocal of the \texttt{pPXF} keyword ``\texttt{REGUL}'') \\
    \texttt{FIXED}            &  Fix stellar kinematics to the results obtained with the PPXF-module? (See correspondent \texttt{pPXF} keyword) \\
    \texttt{NOISE}            &  Pass a constant noise vector to \texttt{pPXF} \\
    \midrule
    \multicolumn{2}{p{6cm}}{\emph{Line Strength Settings}} \vspace{0.1cm} \\
    \texttt{CONV\_COR}        &  Resolution of the index measurement [$\AA$] \\
    \texttt{MC\_LS}           &  Number of Monte-Carlo simulations used to extract errors on the LS indices\\
    \midrule
    \multicolumn{2}{p{6cm}}{\emph{SSP Settings}} \vspace{0.1cm} \\
    \texttt{NWALKER}          &  Number of walkers for the MCMC algorithm (used for the conversion of indices to population properties) \\
    \texttt{NCHAIN}           &  Number of iterations in the MCMC algorithm (used for the conversion of indices to population properties) \\
    \bottomrule
    \bottomrule
    \end{tabular}
    \caption{%
        Overview of all switches and parameters in the main configuration file of the \texttt{GIST} pipeline. This file
        contains one line per galaxy in consideration. 
    }
    \label{tab:configs}
\end{table}

\begin{table}
    \centering
    \begin{tabular}{p{1.7cm}p{6cm}}
    \toprule
    \toprule
    Config-Switch             &  Brief Description \\ 
    \midrule
    i\_line                   & Unique line index \\
    name                      & Name of the emission-line \\
    lambda                    & Restframe wavelength of the emission-line \\
    action                    & Mask, fit, or ignore the emission-line \\
    l-kind                    & Singlet or doublet line? \\
    A\_i                      & Relative amplitude of doublet lines \\
    V\_g/i                    & Guess on the velocity (in restframe) \\
    sig\_g/i                  & Guess on the velocity dispersion or width of the spectral mask \\
    fit-kind                  & Fit lines individually or simultaneously? \\
    AoN                       & Amplitude-over-noise threshold for emission-line subtraction \\
    \bottomrule
    \bottomrule
    \end{tabular}
    \caption{%
        Overview of all switches and parameters in the emission-line configuration file of the \texttt{GIST} pipeline.
        This file contains one line per emission-line in consideration. 
    }
    \label{tab:emissionLinesConfig}
\end{table}

\begin{table}
    \centering
    \begin{tabular}{p{1.7cm}p{6cm}}
    \toprule
    \toprule
    Config-Switch             &  Brief Description \\ 
    \midrule
    b1                        & Minimum wavelength of blue side bandpass \\
    b2                        & Maximum wavelength of blue side bandpass \\
    b3                        & Minimum wavelength of feature bandpass \\
    b4                        & Maximum wavelength of feature bandpass \\
    b5                        & Minimum wavelength of red side bandpass \\
    b6                        & Maximum wavelength of red side bandpass \\
    b7                        & Atomic or molecular index? \\
    names                     & Name of the index \\
    spp                       & Consider this index in the SSP modelling? \\
    origin                    & Comments \\
    \bottomrule
    \bottomrule
    \end{tabular}
    \caption{%
        Overview of all switches and parameters in the line strength configuration file of the \texttt{GIST} pipeline.
        This file contains one line per line strength index in consideration. 
    }
    \label{tab:LSIndicesConfig}
\end{table}

\subsubsection{Preparatory Steps}
\label{subsubsec:preparatorySteps}
The read-in routine of the \texttt{GIST} pipeline performs five simple operations: it reads the already reduced,
science-ready spectral data, shifts all spectra to their rest-frame wavelength while adapting their spectral resolution
correspondingly (according to the configuration parameter \texttt{REDSHIFT}) and shortens them to the wavelength range
defined by the parameters \texttt{LMIN} and \texttt{LMAX}, and rejects spaxels which contain NaN's or have a negative
median flux. Note that those spaxels are not passed to the pipeline, so that the input and output cubes might not have
an identical size. Finally, it computes the signal-to-noise ratio based on the variance spectra obtained during the data
reduction or, if these are not available, exploits the
\texttt{der\_snr}-algorithm\footnote{\url{http://www.stecf.org/software/ASTROsoft/DER\_SNR}; see also
\url{https://www.spacetelescope.org/static/archives/stecfnewsletters/pdf/hst\_stecf\_0042.pdf}} which estimates the
signal-to-noise ratio SNR $= s/n$ using the following equations: 
\begin{eqnarray}
    s   &=&  <F_i> \\
    n   &=&  \dfrac{1.482602}{\sqrt{6}}  < \left| 2F_i - F_{i-2} - F_{i+2} \right| >
\end{eqnarray}
with the signal $s$, the noise $n$, and the flux $F$ at a spectral pixel $i$, where brackets represent the median.
The pipeline is capable of handling data from any IFS instrument, as well as from long-slit or fibre-spectrographs if a
read-in routine that conducts these five operations is provided. In the default version of the pipeline, read-in
routines for the wide- and narrow-field modes of the MUSE spectrograph as well as the V500 and V1200 modes of the
PMAS/PPAK instrument, are included.  These read-in routines can be selected with the configuration switch \texttt{IFU}. 

To run \texttt{pPXF} and \texttt{GandALF}, we first logarithmically rebin the spectra to have constant bins in velocity
(instead of wavelength).  This is implemented in the \texttt{GIST} pipeline by exploiting the log-rebinning function of
\texttt{pPXF} \citep{cappellari2004,cappellari2017}. 

The signal-to-noise ratio of the data is an essential factor in any fitting attempt.  In the specific case of using
\texttt{pPXF} to derive the line of sight velocity distribution (LOSVD) parametrised by the line of sight velocity $v$,
velocity dispersion $\sigma$, and the higher-order Gauss-Hermite moments $h_3$ and $h_4$
\citep{gerhard1993,vanderMarel1993}, higher-order moments critically depend on the signal-to-noise ratio
\citep[see][]{vanderMarel1993,gadotti2005}.  Thus, the pipeline uses the Voronoi binning method of
\citet{cappellari2003} to spatially bin the data. The resulting binned spectra have an approximately constant
signal-to-noise ratio given by the parameter \texttt{TARGET\_SNR}. Note that spaxels which surpass this target
signal-to-noise ratio remain unbinned.  In addition, a minimum signal-to-noise ratio threshold can be applied prior to
the spatial binning. This threshold removes spaxels below the isophote level which has an average signal-to-noise ratio
given by the parameter \texttt{MIN\_SNR} and thus can avoid possible systematic effects in low surface brightness
regions of the data. 

Some IFS data suffer from spatial correlations of the noise between adjacent spaxels. Such correlations can be
introduced, for instance, by the combination of individual observations, spatial interpolation procedures in the
reduction process, or a point-spread-function which significantly surpasses the spatial pixel size of the instrument
\citep[see e.g.][]{garcia-benito2015,sarzi2018}.  If multiple spectra with correlated noise are coadded, the noise of
the resulting stacked spectrum will be underestimated. Depending on the quality of the data, it is of substantial
importance to account for this effect in the Voronoi-binning process. 
\citet{garcia-benito2015} estimate the strength of these spatial correlations $\beta$ by calculating the ratio of the
real noise to the analytically expected noise as a function of the number of spaxels per bin $N$. They find the
following empirical relation (their Eq. 1):
\begin{eqnarray}
    \beta(N) = 1 + \alpha \log N
\end{eqnarray}
where the slope $\alpha$ is a basic proxy for the strength of any, if present, correlations that need to be provided to
the pipeline with the configuration parameter \texttt{COVAR\_VOR}. For instance, in data release 2 of the CALIFA survey
$\alpha$ is found to be 1.06 or 1.07 depending on the setup of the instrument. To account for spatial correlations in
the \texttt{GIST} pipeline, the analytically expected noise is corrected with the above formula during the
Voronoi-binning procedure.  Note that the resulting Voronoi-binning scheme is used throughout the analysis with the
exception of the GANDALF-module that can be run on either the Voronoi bins only or on all spaxels (see
Sect.~\ref{subsubsec:emissionLineAnalysis}). 

In line with the modular structure of the pipeline, spectral template libraries can be selected conveniently with the
configuration parameter \texttt{SSP\_LIB}. While the MILES library \citep{vazdekis2010} is included in this software
package, other libraries are to be provided by the user in distinct directories.  We highlight that the \texttt{GIST}
pipeline allows the user to pick not just a full set of the pre-loaded spectral template libraries, but also an
object-specific subset of such template spectra.  The spectra of the template library are read-in, shortened to conform
with the spectral range of the observed spectra, and oversampled by a factor of two \citep{cappellari2017}. Moreover,
the templates can be normalised to provide mass- or light-weighted results (parameter \texttt{NORM\_TEMP}) and broadened
from their intrinsic, spectral resolution to the one specified by the line-spread-function of the particular instrument. 

For the measurement of stellar population properties via full-spectral fitting with \texttt{pPXF}, the compilation of
the spectral library has further requirements.  More specifically, \texttt{pPXF} necessitates that the templates are
sorted in a three dimensional cube of age, metallicity and $\alpha$-element enhancement, according to the population
properties encoded in the filename of the templates. This is readily implemented in the pipeline for all libraries that
follow the MILES naming convention. Thanks to the modularity of the pipeline, it is straightforward to expand this to
distinct naming conventions, by simply replacing the read-specific function in the SFH-module.

%
\subsubsection{Stellar Kinematics Analysis}
The PPXF-module utilises the penalised pixel-fitting (\texttt{pPXF}) method developed by \citet{cappellari2004} and
advanced in \citet{cappellari2017} (configuration switch \texttt{PPXF}). Its Python implementation derives the
underlying galaxy stellar line-of-sight velocity distribution (LOSVD) parametrised in terms of the line-of-sight
velocity $v$, velocity dispersion $\sigma$, and higher-order Gauss-Hermite moments \citep[switch
\texttt{MOM;}][]{gerhard1993,vanderMarel1993}. Briefly, the method convolves a non-negative linear combination of the
given set of template library spectra with the LOSVD in pixel space by means of a least-squares minimization seeking to
find the set of best-fitting LOSVD parameters and the corresponding weights ascribed to the linear combination of
templates library. 

The regularisation of \texttt{pPXF} is turned off and the default value of the penalization used. Additive (parameter
\texttt{ADEG}) and multiplicative (parameter \texttt{MDEG}) Legendre polynomials can also be included in the fit to
account for any potential deviations in the continuum shape between observed and template spectra or inaccuracies in the
flux calibration. 

In addition to the default functionality of \texttt{pPXF}, extensive Monte-Carlo (MC) simulations can be performed to
compute errors on the extracted quantities (parameter \texttt{MC\_PPXF}). To this end, multiple realisations of the
input spectra are created by adding random noise on the scale of the residual noise to the best fitting result from the
initial \texttt{pPXF} run. Subsequently, the \texttt{pPXF} fit is performed again in the same configuration, but without
the penalization term. The standard deviation of the resulting distribution of the measured quantities is saved as
error. The formal errors on the initial fit, as returned by \texttt{pPXF}, are saved as well. 

Moreover, the PPXF-module computes maps of the $\lambda$ parameter \citep{emsellem2007} which acts as a quantification
of the projected specific angular momentum. Following their Eq.~5 and 6, the parameter is defined as 
\begin{eqnarray}
    \lambda  \equiv  \dfrac{ \left\langle R \left| v \right| \right\rangle}{ \left\langle R \sqrt{v^2 + \sigma^2} \right\rangle }
\end{eqnarray}
and measured in every spatial bin individually as
\begin{eqnarray}
    \lambda  =  \dfrac{ \sum_{i=1}^{N_p} F_i R_i \left| v_i \right| }{ \sum_{i=1}^{N_p} F_i R_i \sqrt{v_i^2 + \sigma_i^2 } }
\end{eqnarray}
with the galactocentric radius $R$ and the flux $F$. Note that the summation is performed over each spaxel in the
corresponding bin, using spaxel-level values for flux and radius and bin-level values for velocity and velocity
dispersion. As this parameter depends on the galactocentric radius, the pipeline only outputs correct values of
$\lambda$ if the coordinates of the centre of the galaxy are passed with the configuration parameter \texttt{ORIGIN}.
Note that the \texttt{GIST} pipeline computes the $\lambda$ parameter in every spatial bin individually, in contrast to
the aperture integrated quantities presented in \citet{emsellem2007}.

\subsubsection{Emission-line Analysis}
\label{subsubsec:emissionLineAnalysis}
For the measurement of gaseous kinematics, emission-line fluxes, and the computation of emission-free continuum spectra
the GANDALF-module facilitates a new Python implementation of the original \texttt{GandALF} procedure by
\citet{sarzi2006} and \citet{jfb2006}, that is released together with this software package (hereafter
\texttt{pyGandALF}; configuration switch \texttt{GANDALF}). There, the individual emission-lines are treated as
additional Gaussian templates and their velocities and velocity dispersions are searched for iteratively.  These
emission-line templates are linearly combined with the set of library spectral templates to solve for the best-fitting
emission-line gaseous kinematics (i.e. velocity and velocity dispersion), their strengths (i.e. line amplitudes), and
weights on the linear combination of the spectral template library.  To this extent, such a procedure adequately
accounts for both the gaseous emission-lines and the underlying stellar continuum present in the spectra of some
galaxies \citep{sarzi2006}. 

Following the original implementation of \texttt{GandALF}, both the GANDALF-module and \texttt{pyGandALF} only fit
emission-lines which are specified by the user in the emission-line configuration file (see
Table~\ref{tab:emissionLinesConfig}). In particular, this file states the wavelength of the line and whether the line is
a singlet or doublet, and whether different sets of lines are to share the same kinematics or not. If specified in the
emission-line configuration file, multiple emission-line components can be fitted to a single emission feature, in order
to reproduce non-Gaussian line profiles. For instance, in the presence of an outflow, a narrow emission-line might be
located on top of a very broad emission feature. In this case, the broad and the narrow component can be fitted
independently, thus obtaining an improved fit and results on the two physically distinct components. 

Initial guesses on the velocity (relative to the redshift given with the parameter \texttt{REDSHIFT}) and velocity
dispersion are to be provided by the user in the emission-line configuration file.  \texttt{pyGandALF} can compensate
differences in the continuum from the galaxy and the templates by exploiting a two-component reddening correction
(parameter \texttt{REDDENING}).  As detailed in \citet{oh2011} and shown also in \citet{sarzi2018} this allows to
construct maps for classical ``screen''-like extinction affecting the entire galaxy spectrum in addition to maps for the
reddening impacting only on the emission-line regions. Alternatively, if the \texttt{REDDENING} option is not used, the
same order of multiplicative Legendre polynomials as for the PPXF-module is applied in \texttt{pyGandALF}. Additive
Legendre polynomials are not used, regardless of what is set in the configuration parameter \texttt{ADEG}.  In addition,
the GANDALF-module is capable of correcting the spectra for Galactic extinction in the direction of the target. To this
end, Galactic extinction values \citep[e.g.  from][]{schlegel1998} are passed with the configuration parameter
\texttt{EBmV} and the spectra are de-reddened prior to the emission-line analysis by applying the model of
\citet{calzetti2000}. 

In addition to its measurement of gaseous kinematics, the GANDALF-module also computes continuum spectra with subtracted
emission-lines (hereafter \emph{emission-subtracted spectra}). A notable complication of this step is the determination
of a threshold above which a line detection is considered significant. To this end, \citet{sarzi2006} introduce the
amplitude-to-residual noise ratio (A/rN) which quantifies how much the amplitude of an emission-line surpasses the
residual noise level of the spectrum, which can be used to justify this choice. In the default setup of the
\texttt{GIST} pipeline, the best-fitting emission-line templates are subtracted from the observed spectra, if this A/rN
ratio (using the A/rN value of the main line for doublets of multiplets) exceeds the A/rN threshold defined in the
emission-line configuration file (see Table~\ref{tab:emissionLinesConfig}).
The impact of pegging lines together on the minimum A/rN values to robustly detect lines was initially discussed in
\citet{sarzi2006} in the case of SAURON observations, but a more detailed study is needed to robustly address this issue
for more complicated line sets as observed within the MUSE wavelength range and for different kind of emission-line
systems.
Thanks to the modularity of the code, it is straightforward to implement different methods of the emission-line
subtraction, by simply changing one function in the GANDALF-module.  Note that we are currently in the process of
developing more sophisticated criteria for the selection of such a detection threshold and a more advanced treatment of
the emission-line subtraction problem could be released in a foreseeable pipeline release.

\subsubsection{Non-parametric Star Formation Histories}
The SFH-module of the \texttt{GIST} pipeline exploits \texttt{pPXF} to estimate some of the underlying
stellar-population properties and their associated non-parametric star formation histories via full-spectral fitting
(configuration switch \texttt{SFH}). Essentially, the code finds a linear combination of spectral templates that
resembles the observed spectrum.  Under such an assumption the stellar-population properties and the corresponding
star-formation histories are derived on the basis of the linear weights ascribed to the chosen set of template spectra.
However, recovering this information from the observational data is an inverse, ill-conditioned problem. Small
variations in the initial data can translate to large variations in the solution.  Thus, regularisation is used to
obtain a more physically motivated combination of the spectral template library. In particular, of all solutions that
are equally consistent with the observational data, the regularisation algorithm returns the smoothest solution.  We
refer the interested reader to Sect. 3.5 of \citet{cappellari2017} for a detailed review of the stellar population
analysis with \texttt{pPXF}. 

The regularisation parameter, corresponding to the \texttt{pPXF} keyword \texttt{REGUL}, acts as a proxy for the
strength of the regularisation and is passed to the pipeline with the configuration parameter \texttt{REGUL\_ERR}, a
value which is the reciprocal of the \texttt{pPXF} keyword \texttt{REGUL}.  The given value is subsequently used for all
spectra in the cube.  Note that the choice of this regularisation parameter could have a substantial effect on some of
the recovered population properties.

The SFH-module uses emission-subtracted spectra, as produced by the GANDALF-module, for the derivation of population
properties.  During this process, the parameters representing the stellar kinematics can be extracted simultaneously
with the population properties or kept fixed to those obtained with the PPXF-module (configuration switch
\texttt{FIXED}, see also correspondent \texttt{pPXF} keyword).  Deviations in the continuum shape between observed and
template spectra can be compensated by including a multiplicative Legendre polynomial in the fit (paramter
\texttt{MDEG}; same as in the PPXF-module). As the use of an additive Legendre polynomial could significantly alter the
resulting population properties, those cannot be used in the SFH-module, regardless of what is set in the configuration
parameter \texttt{ADEG}.

Note that the estimation of the errors on the population properties, as derived from regularised full-spectral fitting,
is somewhat ambiguous and thus we do not attempt to provide such a procedure in this software package. An ensemble of
different approaches have been previously undertaken \citep[see e.g.][]{gadotti2019,pinna2019} and therefore we advise
the user to find the most suitable procedure depending on their scientific context.

\subsubsection{Absorption-Line Strength Analysis}
Complementary to the derivation of stellar population properties via full-spectral fitting, a wide range of studies has
made use of absorption-line strength indices to infer information about the stellar content of galaxies
\citep[e.g.][]{trager1998,peletier2007,kuntschner2010,mcdermid2015}.The fundamental steps of the line strength
measurement are the following: a main absorption feature bandpass is encompassed by two side bandpasses to the red and
blue. The side bandpasses act as a proxy for the stellar continuum. To this end, the mean flux in both side bandpasses
is computed and their central points connected by a straight line. The flux difference between the observed spectrum and
this straight line calculated within the wavelength range of the feature bandpass represents the absorption-line
strength index \citep[see e.g.][]{faber1985,kuntschner2006}. 

The LS-module of the \texttt{GIST} pipeline makes it possible to measure absorption line strengths in the \emph{Line
Index System} \citep[LIS;][]{vazdekis2010} by adopting a Python translation of the routine used by
\citet{kuntschner2006} (configuration switch \texttt{LINE\_STRENGTH}). The wavelengths of the feature and continuum
bandpasses are defined by the user in a dedicated configuration file (see Table~\ref{tab:LSIndicesConfig}). This
configuration file further states whether the index will be computed as equivalent width in Angstroms or in magnitudes
(as commonly used for atomic and molecular absorption features, respectively). 

In preparation of this measurement, the LS-module uses emission-subtracted spectra, translates those from velocity space
back to constant bins in wavelength, and convolves them to the spectral resolution of the measurement (parameter
\texttt{CONV\_COR} in the main configuration file). Note that this convolution accounts for both the instrumental and
stellar velocity dispersion. The measured stellar radial velocity is considered to assure the correct placement of the
line strength bandpasses.  Errors are estimated by means of Monte-Carlo simulations, considering errors on the stellar
radial velocity and the variance spectra (parameter \texttt{MC\_LS} in the main configuration file). 

As a second independent step, the measured line strengths can be matched to SSP-equivalent properties.  The main
ingredient is a model file provided together with the selected spectral library (parameter \texttt{SSP\_LIB} in the main
configuration file) at the corresponding measurement resolution (parameter \texttt{CONV\_COR}) which connects line
strength indices to population properties. Taking into account the errors in the indices, radial velocities and variance
spectra, a Markov chain Monte Carlo (MCMC) algorithm \citep[][]{emcee2013} searches for the best SSP to match the
derived set of indices \citep[parameters \texttt{NWALKER} and \texttt{NCHAIN}; see][for further
details]{martin-navarro2018}.

\section{Application to MUSE data}
\label{sec:application}
In this section we aim to illustrate the capabilities of the pipeline by applying it to real data.  We present the full
set of high-level data products, in particular the resulting stellar kinematics and emission-line analysis, as well as
absorption line strength indices and stellar population properties for one galaxy. We highlight that all figures
presented in this section, with the exception of Fig.~\ref{fig:intensities}, are automatically generated by the
pipeline.  Moreover, the pipeline outputs immediately allow the user to produce further plots, for instance BPT diagrams
and maps of electron densities. 

We exploit observations of the galaxy NGC\,1433 obtained with the MUSE integral-field spectrograph as part of the TIMER
survey. All galaxies in the TIMER sample are barred and exhibit a variety of nuclear structures, such as nuclear rings,
inner discs and primary/secondary bars, as well as nuclear spiral arms \citep[see
e.g.][]{mendez-abreu2019,deLorenzoCaceres2019}. The main scientific goal of the project is to study the star-formation
histories of these nuclear structures. More specifically, this allows to determine the cosmic time at which the bar
initially formed and thus constrain the epoch at which the galaxy's disc became dynamically mature
\citep[see][]{gadotti2015,gadotti2019}.

We choose the galaxy NGC\,1433 as it is a typical example of a spiral galaxy with various structural components, which
we expect to be clearly distinguishable in their kinematic and stellar population properties. In addition, this galaxy
exhibits a variety of ionized-gas emission, without showing significant emission from an AGN or suffering from severe
dust contamination. Therefore, NGC\,1433 represents a good example to demonstrate the capabilities of the \texttt{GIST}
pipeline.

\subsection{Observations and Data Reduction}
\label{subsec:observation}
The observations of NGC\,1433 were performed in ESO Period 97 in August and October 2016 with an average seeing of 0.8 -
0.9\,arcsec. The MUSE field-of-view (FoV) covers approximately 1\,arcmin $\times$ 1\,arcmin with a spatial sampling of
0.2\,arcsec $\times$ 0.2\,arcsec. It covers a spectral range from 4750 to 9350\,$\AA$ with a spectral pixel size of
1.25\,$\AA$ and mean spectral resolution of 2.65\,$\AA$.  The data reduction is based on the version 1.6 of the
dedicated ESO MUSE data-reduction pipeline \citep{weilbacher2012}.  In summary, bias, flat-fielding and illumination
corrections are applied and the exposures calibrated in flux and wavelength. Telluric features and signatures from the
sky background are removed and exposures are registered astrometrically. For details on observations and data reduction
we refer the reader to \cite{gadotti2019}. 

NGC\,1433 is a barred, early-type spiral with an inner ring, plume, nuclear ring-lens and nuclear bar, as well as an
outer pseudo-ring \citep[classified as \emph{(R\textquotesingle $_1$)SB(r,p,nrl,nb)a} by ][]{buta2015}.  Its inclination
is 34 degrees based on the 25.5\,AB\,mag\,arcsec$^{-2}$ isophote at 3.6\,$\mu$m and it has a stellar mass of $2.0 \times
10^{10} \mathrm{M_{\odot}}$.  The mean redshift-independent distance from the NASA Extragalactic Database
\footnote{\url{http://ned.ipac.caltech.edu/}} is 10.0\,Mpc \citep[see also][and references therein]{gadotti2019}. 

\begin{figure}
    \includegraphics[width=0.49\hsize]{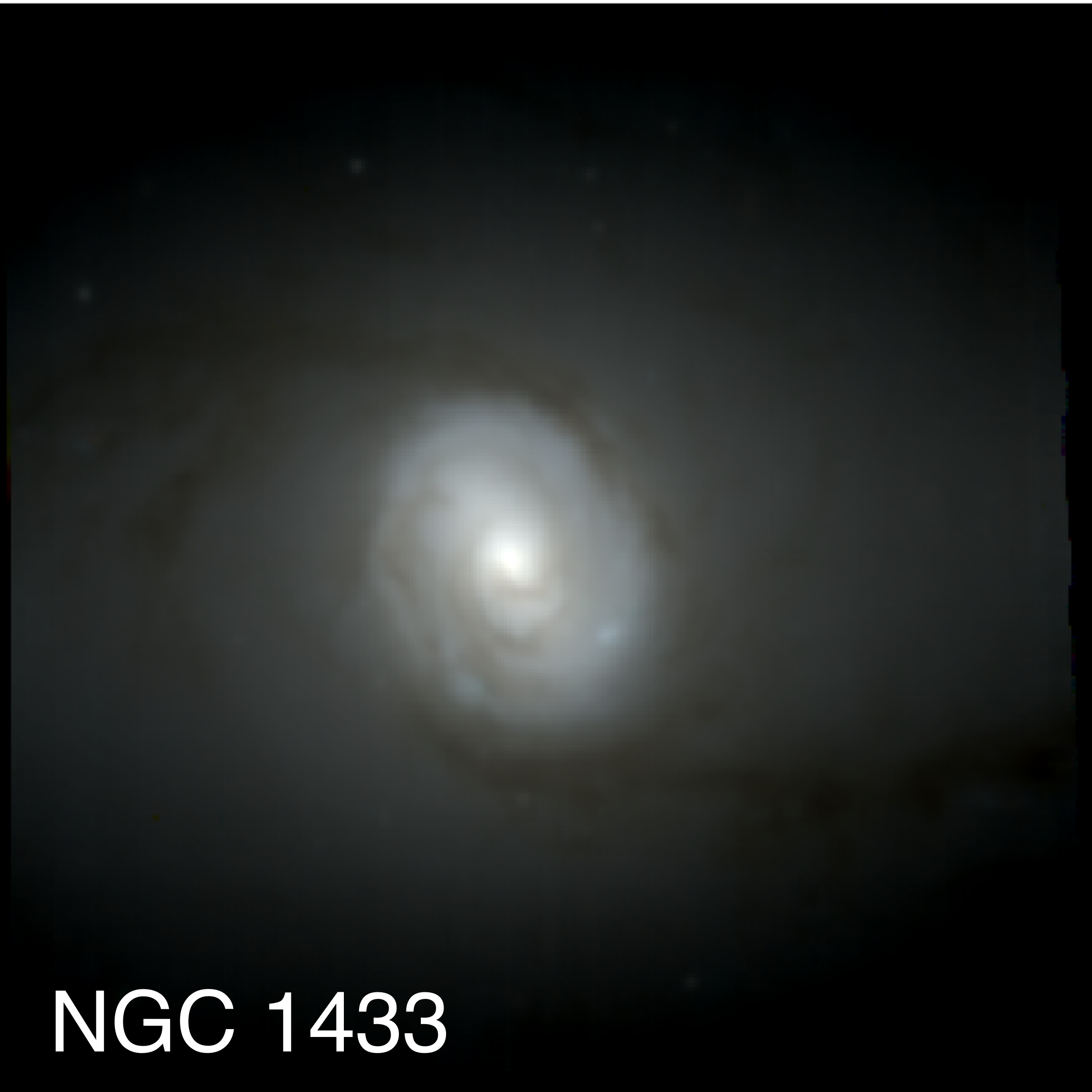}
    \includegraphics[width=0.49\hsize]{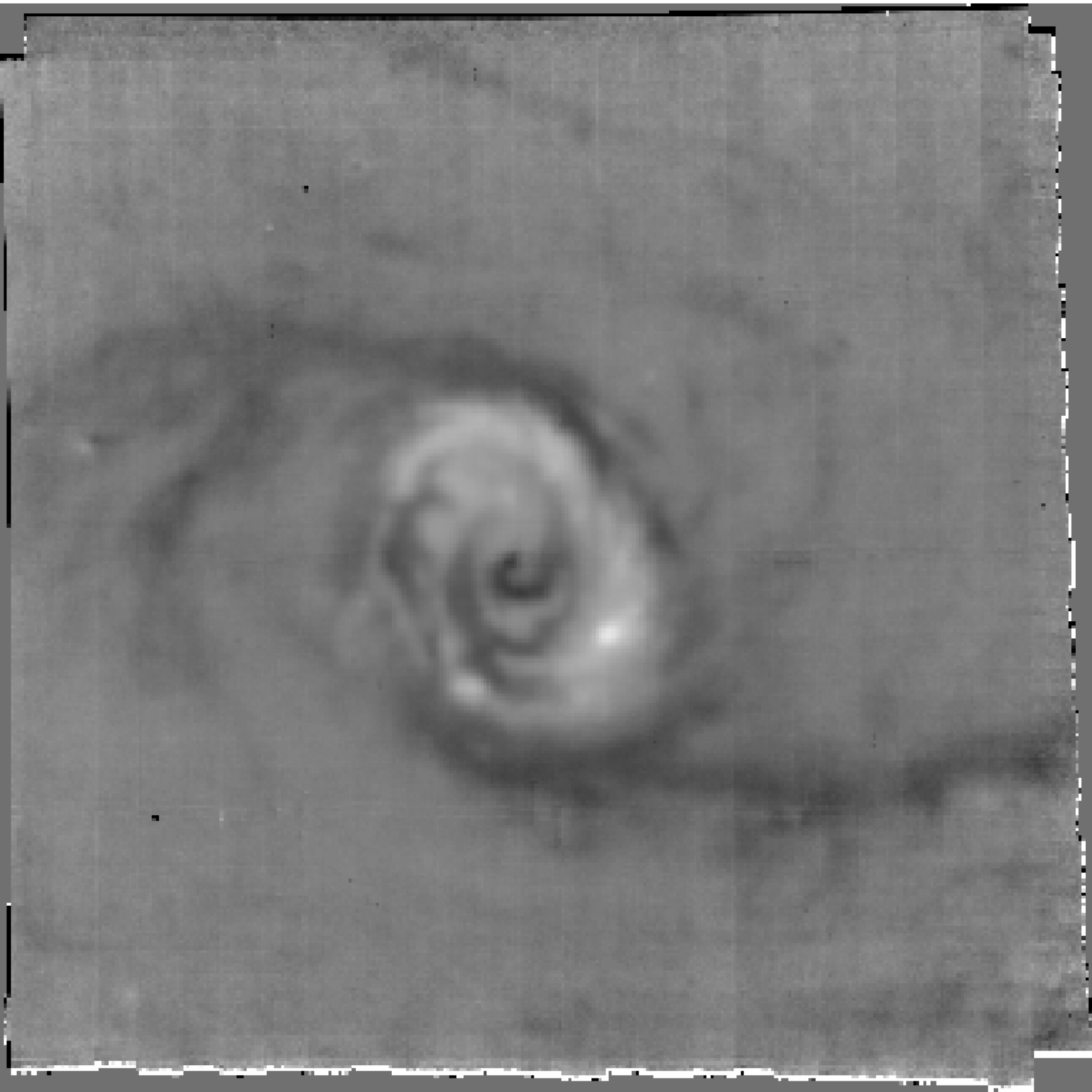}
    \caption{%
        Colour composite (left-hand panel) and colour map (right-hand panel) of NGC\,1433. Colour composites are built
        by collapsing the data cube in a blue (4750 to 6000\,$\AA$), green (6000 to 7000\,$\AA$), and red (7000 to
        9000\,$\AA$) wavelength band and combining the resulting images. Colour maps are obtained by subtracting the red
        from the blue image. 
        }
    \label{fig:intensities}
\end{figure}

In Fig.~\ref{fig:intensities} we further illustrate the most prominent structures in the MUSE FoV by presenting a colour
composite (left-hand panel) and colour map (right-hand panel).  The colour composite highlights a star-forming inner
disc; however, in contrast to previous studies \citep[e.g.][]{erwin2004,buta2015} we do not find evidence for a nuclear
bar in the MUSE reconstructed intensities.  The colour map features two strong dust lanes along the leading edges of the
main bar. In fact, dust lanes along the leading edges of the bar are expected from theoretical work
\citep{athanassoula1992}, and in the case of NCG\,1433 end at the inner disc \citep[see also][]{sormani2018}. 

\subsection{Data Analysis Setup}
\label{subsec:dataAnalysisSetup}
We perform the data analysis with the pipeline configuration setup discussed in Sect.~\ref{sec:code}. Throughout this
analysis, we spatially bin the data to an approximately constant signal-to-noise ratio of 40 per bin. This
signal-to-noise level has been widely used in literature, as it provides a good compromise between accuracy of the
extracted quantities and spatial resolution \citep[see e.g.][]{vanderMarel1993}. Therewith it is a suitable choice for
this demonstration of the \texttt{GIST} pipeline. Spaxels which surpass this ratio remain unbinned in the analysis,
while all spaxels below the isophote level which has an average signal-to-noise ratio of 3 are excluded, in order to
avoid systematic effects at low surface brightness.  Note that the emission-line analysis is performed on a
spaxel-by-spaxel basis instead, also including spaxels below this signal-to-noise threshold. 

The line-spread-function of the MUSE observations varies with wavelength. To correct for this effect, we adopt the
udf-10 prescription obtained by \citet{bacon2017}. In particular, we broaden all template spectra to the one obtained in
their work prior to performing any fits. 

For the measurement of stellar kinematics and emission-line properties, we use the E-MILES model library
\citep{vazdekis2015}. The library consists of SSP spectra, covering a large parameter space in age and metallicity at a
spectral resolution of $\approx 2.51\,\AA$ in the relevant wavelength range \citep{jfb2011}. In this analysis, we
exploit the rest-frame wavelength range from 4800 to 8950\,$\AA$, in order to maximise the available information.  Any
emission-lines in the considered wavelength range are masked. When extracting stellar kinematics with \texttt{pPXF}, we
include a low-order, multiplicative Legendre polynomial in the fit, to correct for any small differences between
observed and template spectra.  In contrast, the emission-line analysis exploits a two-component reddening correction
instead of Legendre polynomials. 

In order to infer the stellar population properties, we restrict the wavelength range to 4800 to $5500\AA$. This allows
the use of the MILES SSP model library \citep{vazdekis2010}, which covers a shorter wavelength range, but provides
information on the $\alpha$-enhancement of the stellar populations.  The model library covers ages from 0.03 to
14.00\,Gyr, metallicities of -2.27 < [M/H] < +0.40, and $\alpha$-enhancement values [$\alpha$/Fe] of either 0.00 and
0.40. The models are based on the BaSTI isochrones
\citep{pietrinferni2004,pietrinferni2006,pietrinferni2009,pietrinferni2013} and a revised Kroupa initial mass function
\citep{kroupa2001}.  The measurement of the stellar population content is based on the Voronoi-binned,
emission-subtracted spectra, as returned by the GANDALF-module. 

The inference of the non-parametric star-formation histories is performed by a regularised run of \texttt{pPXF}. Stellar
kinematics are kept fixed to those obtained from the unregularised run of \texttt{pPXF}. The strength of the
regularisation is determined following the criterion introduced by \citet{press1992} and applied, for instance, in
\citet{mcdermid2015}. In particular, the maximum value of the regularisation corresponds to the fit for which the
$\chi^2$ of the best-fitting solution increases from the $\chi^2$ of the unregularised solution by $\Delta \chi^2 =
\sqrt{2N_{\text{pix}}}$ with $N_{\text{pix}}$ being the number of fitted spectral pixels. This regularisation parameter
is determined on the bin with the highest signal-to-noise ratio and subsequently applied to the entire cube. 

We measure absorption-line strength indices in the LIS-8.4 system \citep{vazdekis2010}. To this end, all spectra are
convolved to a spectral resolution of 8.4\,$\AA$, taking into account both instrumental and local stellar velocity
dispersion. Errors on these indices are estimated through 30 Monte-Carlo simulations. Subsequently, the MILES SSP model
which best predicts the observed combination of the H$\beta_{\circ}$ \citep{cervantes2009}, Fe5015 \citep{worthey1994},
Mgb \citep{burstein1984}, Fe5270 \citep{burstein1984}, and Fe5335 \citep{worthey1994} indices, as used by previous
studies to determine SSP properties, is determined by means of the previously mentioned MCMC fitting technique
\citep{martin-navarro2018}.

\subsection{Stellar Kinematics}
\label{subsec:stellarKinematics}
\begin{figure}
    \includegraphics[width=\hsize]{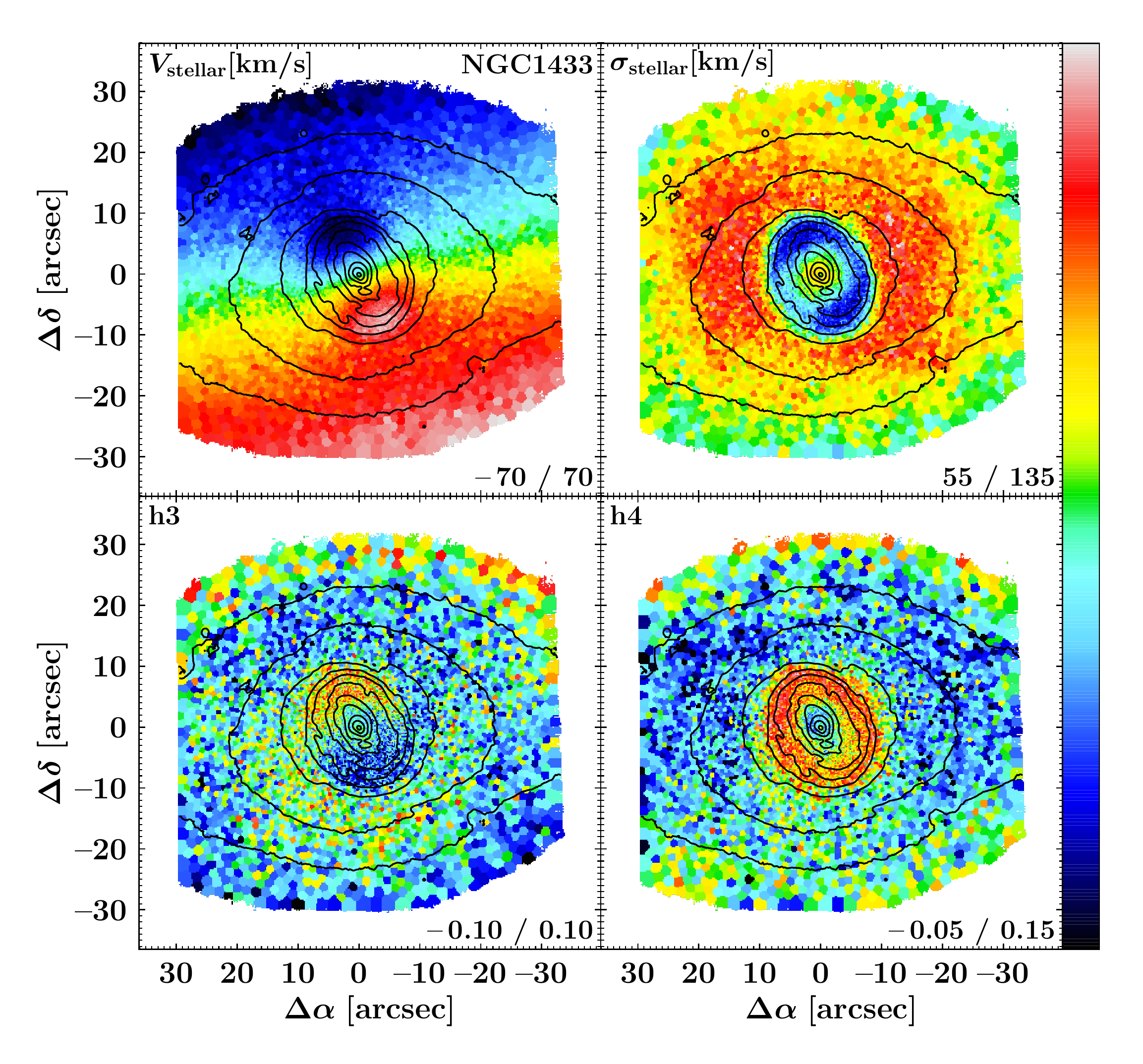}
    \includegraphics[width=\hsize]{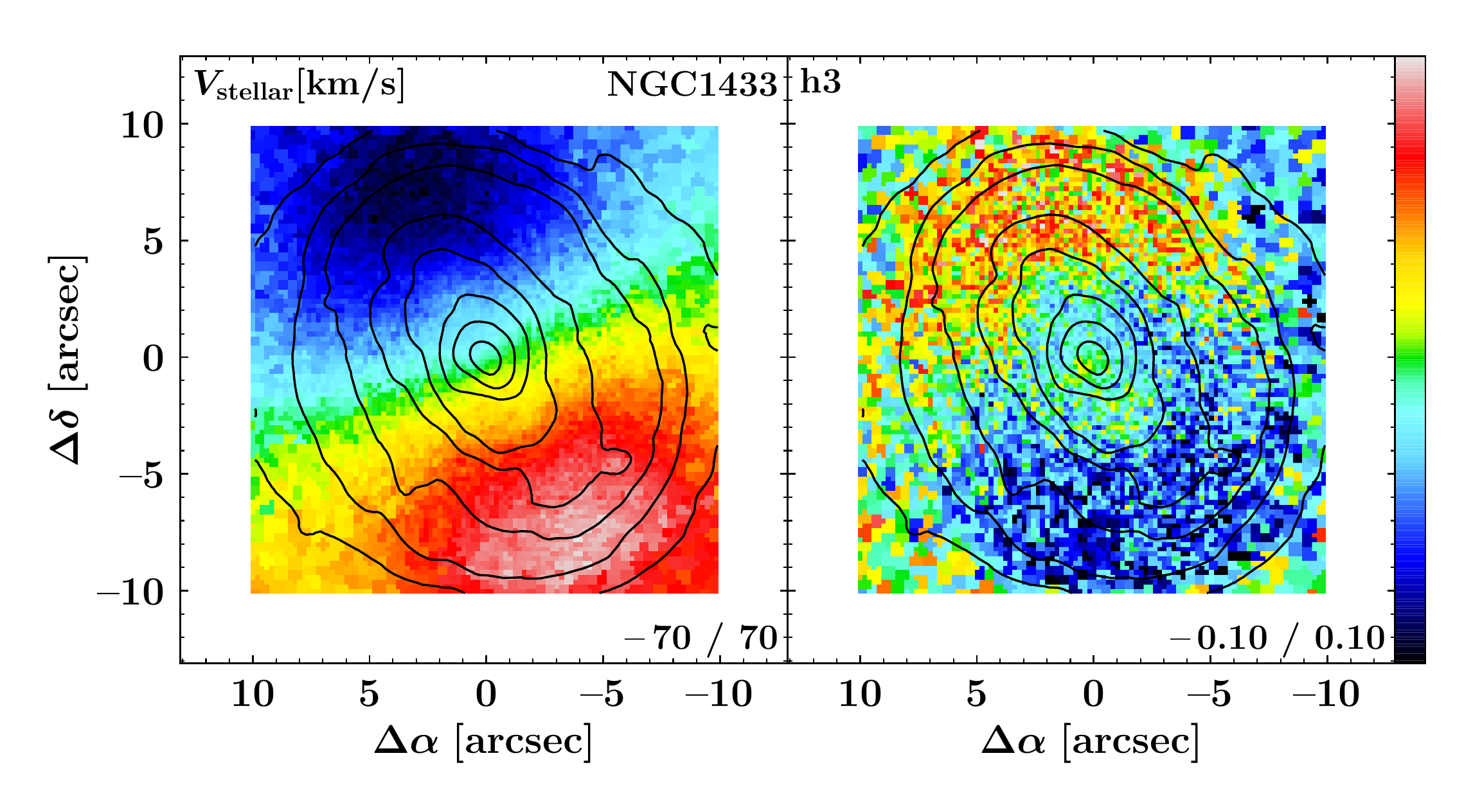}
    \caption{%
        The upper group of panels present the line-of-sight velocity ($V_{\mathrm{stellar}}$), velocity dispersion
        ($\sigma_{\mathrm{stellar}}$), and higher-order moments $\mathrm{h_3}$ and $\mathrm{h_4}$ of the stellar
        component of NGC\,1433, as indicated in the upper left of the panels. 
        The lower group of panels highlights the spatial distribution of line-of-sight velocities (left-hand panel) and
        h$_3$ moment (right-hand panel) in the innermost part of the galaxy.  
        The limits of the colourbar are stated in the lower right of each panel. Isophotes are based on reconstructed
        intensities from the MUSE cube and displayed in steps of 0.5 magnitudes. North is up, east is to the left. 
    }
    \label{fig:stellarKinematicsMap}
\end{figure}

\begin{figure}
    \includegraphics[width=\hsize]{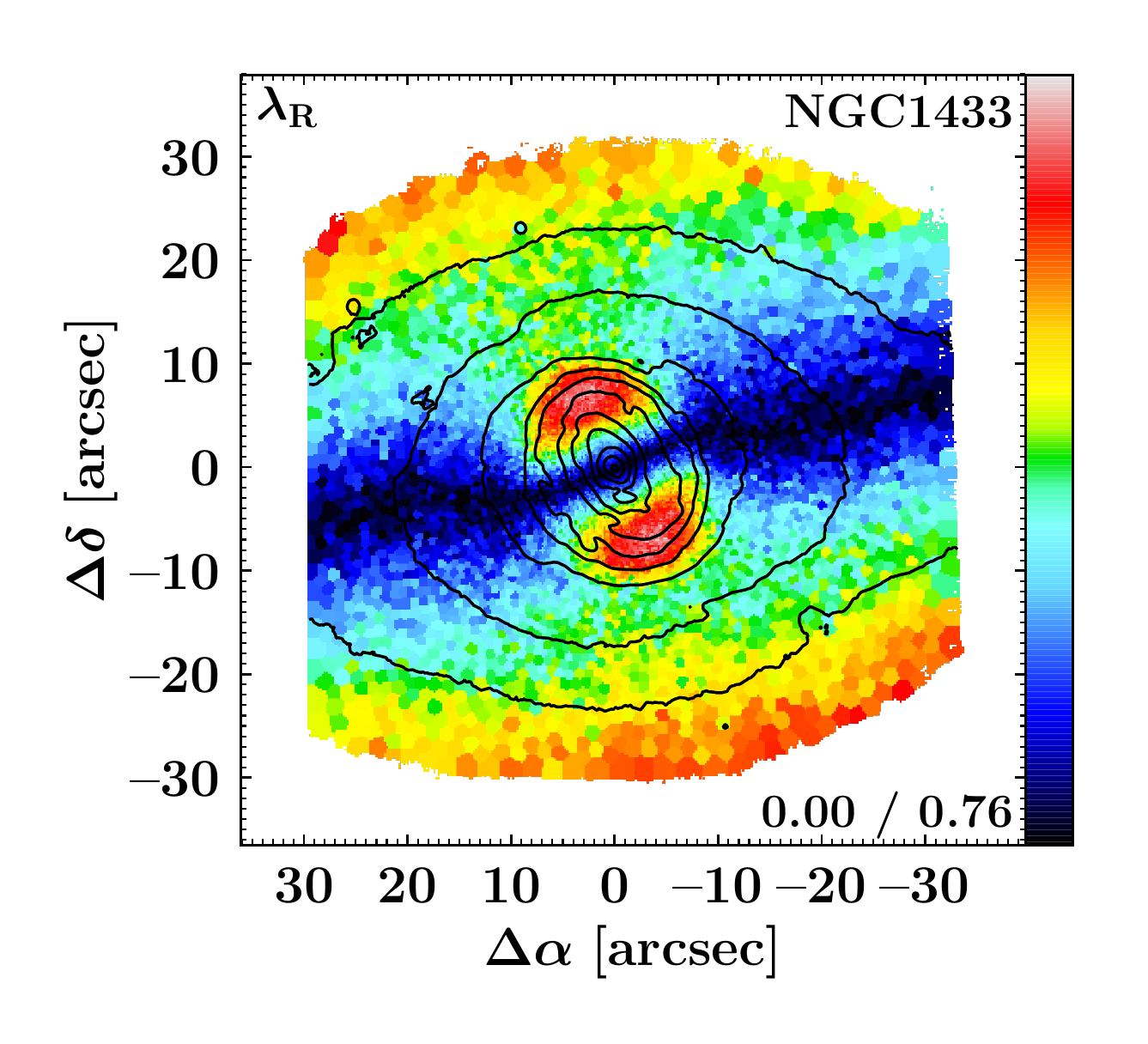}
    \caption{%
        Map of the projected specific angular momentum $\lambda$ of NGC\,1433.  The limits of the colourbar are stated
        in the lower right. Isophotes are based on reconstructed intensities from the MUSE cube and displayed in steps
        of 0.5 magnitudes. North is up, east is to the left.  
    }
    \label{fig:lambdaRMap}
\end{figure}

We present the derived maps of stellar kinematics in the upper group of panels in Fig. \ref{fig:stellarKinematicsMap}
and projected specific stellar angular momentum $\lambda$ in Fig. \ref{fig:lambdaRMap}. The typical formal errors on the
fit are 4.1~km\,s$^{-1}$ in velocity, 5.7~km\,s$^{-1}$ in velocity dispersion, 0.03 in h$_3$, and 0.04 in h$_4$. Various
distinct kinematic features are immediately evident, thanks to the outstanding spatial resolution of the MUSE
observations.  A rapidly rotating inner disc with a radial extent of approximately 10~arcsec is detected in the
line-of-sight velocity map. This inner disc appears even more pronounced in the map of the $\lambda$ parameter.  The map
of the stellar velocity dispersion shows a prominent ring of low velocity dispersion which coincides with the rapidly
rotating inner disc found in the line-of-sight velocity map. This ring is surrounded by a region of elevated velocity
dispersion.  Within the ring, at radii smaller than 3-4~arcsec, the velocity dispersion increases again towards the
centre. 

Correlations between the higher-order Gauss-Hermite moment h$_3$ and the line-of-sight velocity allow to infer
information on the eccentricity of the underlying stellar orbits. For instance, an \emph{anti-correlation} between
line-of-sight velocity and h$_3$ indicates near-circular orbits, as they are typically found in regularly rotating
stellar systems, such as stellar discs \citep[][]{bender1994,bureau2005}. Such an anti-correlation is evident in the
region of the inner disc, in particular at radii between 4 and 10~arcsec. In addition, such an anti-correlation is found
at radii above 25~arcsec, which corresponds to the main disc of the galaxy.  In contrast, a \emph{correlation} between
line-of-sight velocity and h$_3$ is a signature of orbits with high eccentricity. We find such a correlation at
intermediate radii between the inner and main stellar disc of NGC\,1433. In fact, this region might be dominated by the
eccentric x$_1$ orbits of the bar. Interestingly, a similar correlation between line-of-sight velocity and h$_3$ is
found within the nuclear ring of the galaxy, at radii smaller than 3-4~arcsec (see lower group of panels in
Fig.~\ref{fig:stellarKinematicsMap}). This finding is particularly interesting in the context of the potential existence
of an inner bar in NGC\,1433. However, we note that previous studies \citep[see
e.g.][]{erwin2004,buta2015,deLorenzoCaceres2019} remain inconclusive on the existence of an inner bar in this galaxy. 

Finally, high values of the higher-order moment h$_4$ indicate a superposition of structures with different LOSVDs
\citep[see e.g.][and references therein]{bender1994}. For NGC\,1433 this is only the case in the nuclear ring, while in
other parts of the galaxy, in particular in the centre, no elevated h$_4$ values are found. 
  

\subsection{Emission-line Analysis}
\label{subsec:emissionLineAnalysis}

\subsubsection{Emission-line Kinematics and Fluxes}
\label{subsubsec:emissionLineKinematics}
\begin{figure}
    \includegraphics[width=0.49\hsize, trim=0.5cm 0.7cm 0.5cm 0.7cm, clip=true]{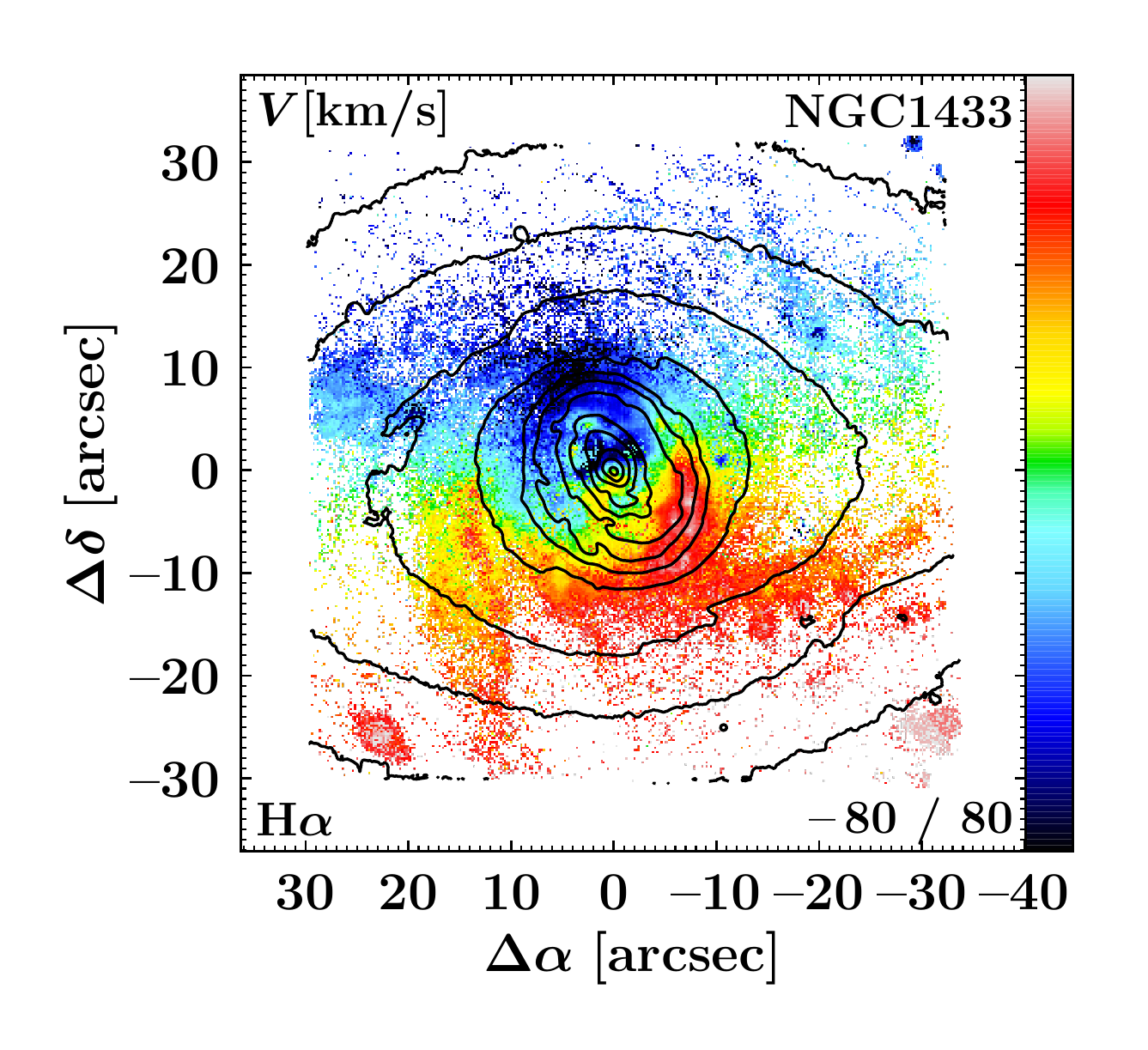}
    \includegraphics[width=0.49\hsize, trim=0.5cm 0.7cm 0.5cm 0.7cm, clip=true]{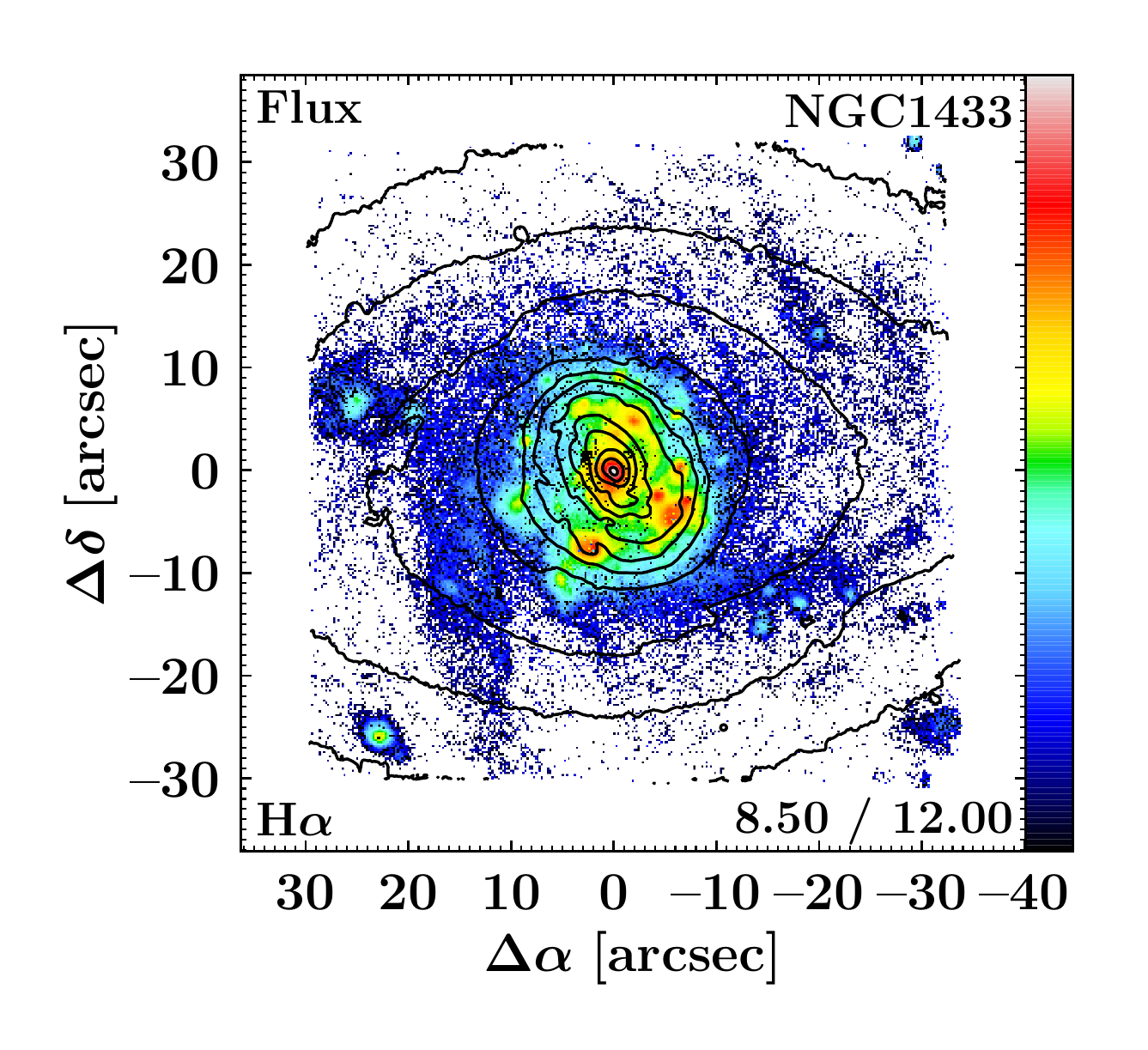}\\
    \includegraphics[width=0.49\hsize, trim=0.5cm 0.7cm 0.5cm 0.7cm, clip=true]{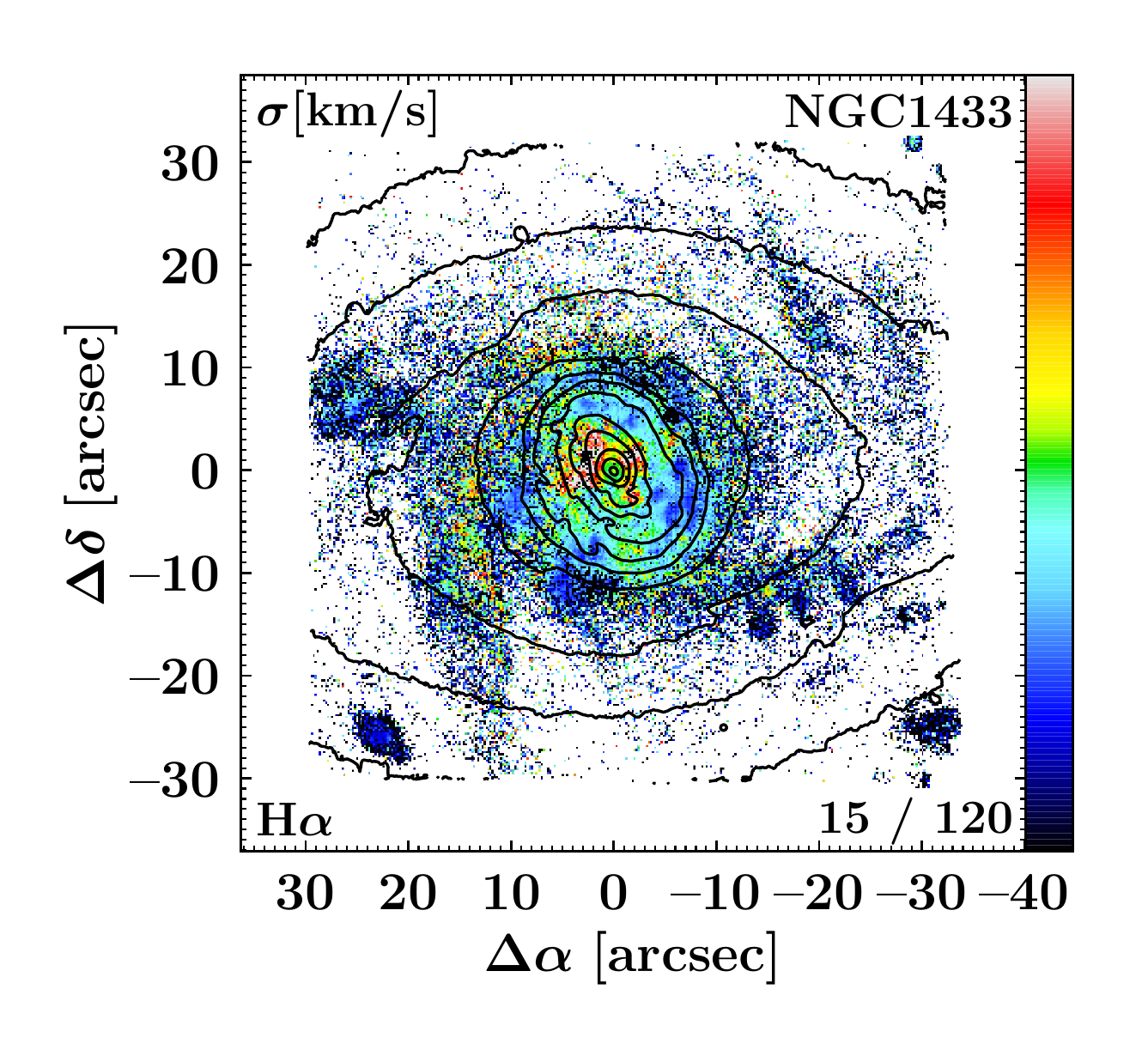}
    \includegraphics[width=0.49\hsize, trim=0.5cm 0.7cm 0.5cm 0.7cm, clip=true]{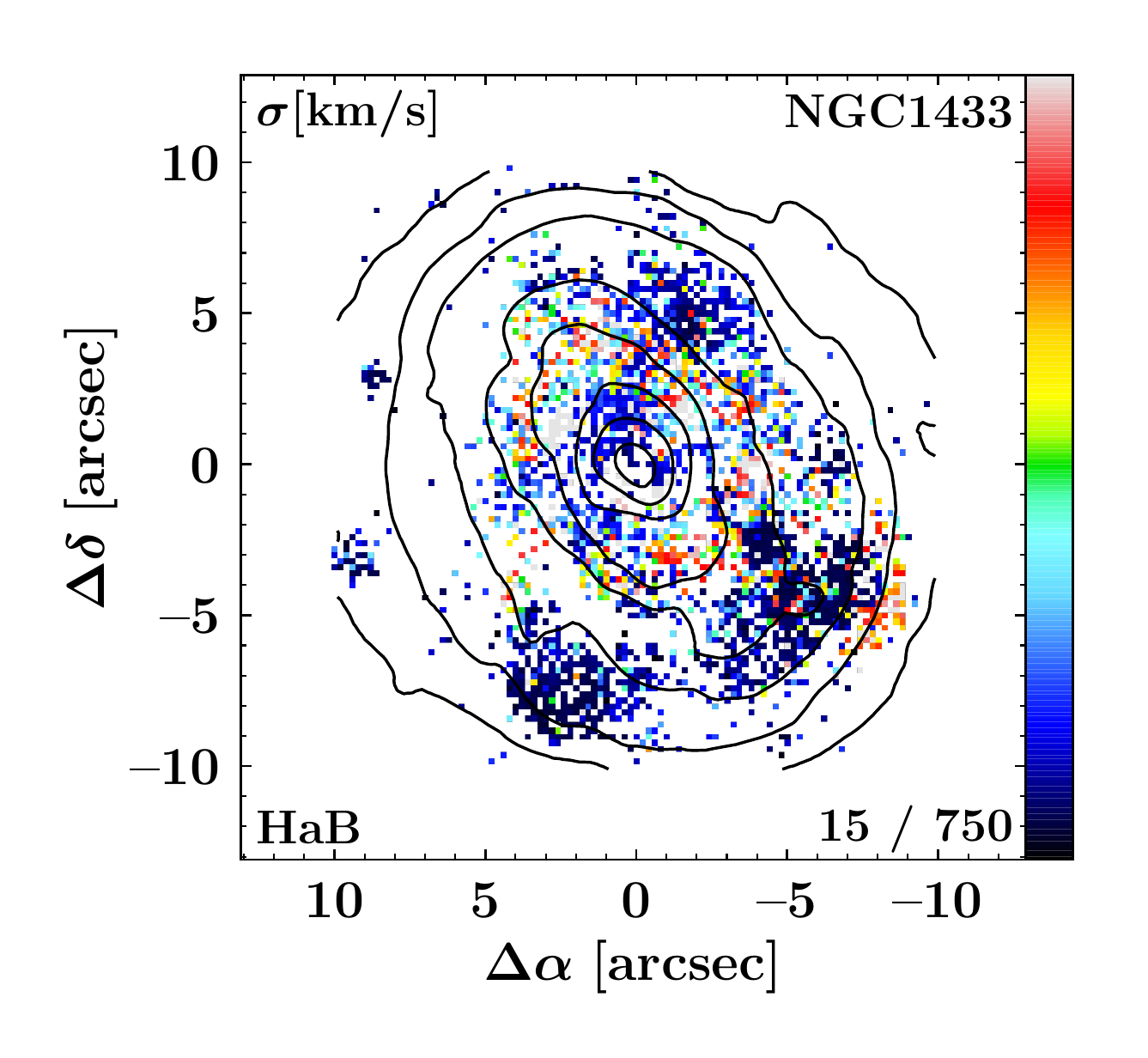}
    \caption{%
        Maps of the H$\alpha$ line-of-sight velocity (upper left-hand panel), velocity dispersion (lower left-hand
        panel) and flux (upper right-hand panel, in arbitrary units with logarithmic scale) of the primary component in
        the emission-line fit of NGC\,1433. The velocity dispersion of the secondary H$\alpha$ component is displayed in
        the lower right-hand panel. Note that this panel shows a different spatial scale, in order to highlight the
        innermost region of the galaxy. Only spaxels in which the H$\alpha$ line exceeds an A/rN ratio of 4 are
        displayed in this figure. The limits of the colourbar are stated in the lower right corner of each panel.
        Isophotes are based on reconstructed intensities from the MUSE cube and displayed in steps of 0.5 magnitudes.
        North is up, east is to the left.  
    }
    \label{fig:Halpha}
\end{figure}

The upper right-hand panel of Fig.~\ref{fig:Halpha} shows the spatial distribution of measured H$\alpha$ fluxes. A large
amount of H$\alpha$ emission, especially in the inner disc, is obvious. While the emission in the inner disc appears
patchy, the highest H$\alpha$ fluxes are found in the centre of the galaxy.  In the upper left-hand and lower left-hand
panels of Fig.~\ref{fig:Halpha}, we present line-of-sight velocity and velocity dispersion maps of the emission-line fit
of H$\alpha$. The line-of-sight velocity map clearly shows H$\alpha$ rotation along the major axis of the galaxy.
However, significant departures from regular rotation, in particular in the region of the inner disc, are obvious. The
spatial distribution of the H$\alpha$ velocity dispersion reveals several areas of low sigma, in particular in the
nuclear ring. These spatially coincide with areas of high H$\alpha$ fluxes, where star formation is proceeding. Close to
the centre of the galaxy, three regions of elevated velocity dispersion are evident. 

The \texttt{pyGandALF} emission-line fit included a secondary component for the H$\alpha$ emission wherever necessary.
In particular, a visual inspection of the one-component fits revealed that a secondary component was necessary in few
regions, even though it is fainter than the primary component. Where such a component is not necessary,
\texttt{pyGandALF} does not recover this component. Note that only few spaxels, mostly located in the inner disc,
include such a secondary emission-line component. The map of the velocity dispersion of this secondary component is
presented in the lower right-hand panel of Fig.~\ref{fig:Halpha}.  In three individual regions in the inner disc, the
secondary component is included as additional, low-dispersion component. Those regions have already been identified as
low-dispersion regions in the primary component of the fit.  Interestingly, in the inner regions of the inner disc, the
secondary component shows high velocity dispersion of several hundred km\,s$^{-1}$. Note that while the secondary
component improved the H$\alpha$ emission-line fit in the shown regions, a detailed investigation on its origin, in
particular whether its origin is physical or not, is beyond the scope of this study. 

\subsection{Line Strength Analysis}
\label{subsec:lineStrengthAnalysis}
\begin{figure*}
    \includegraphics[width=\textwidth]{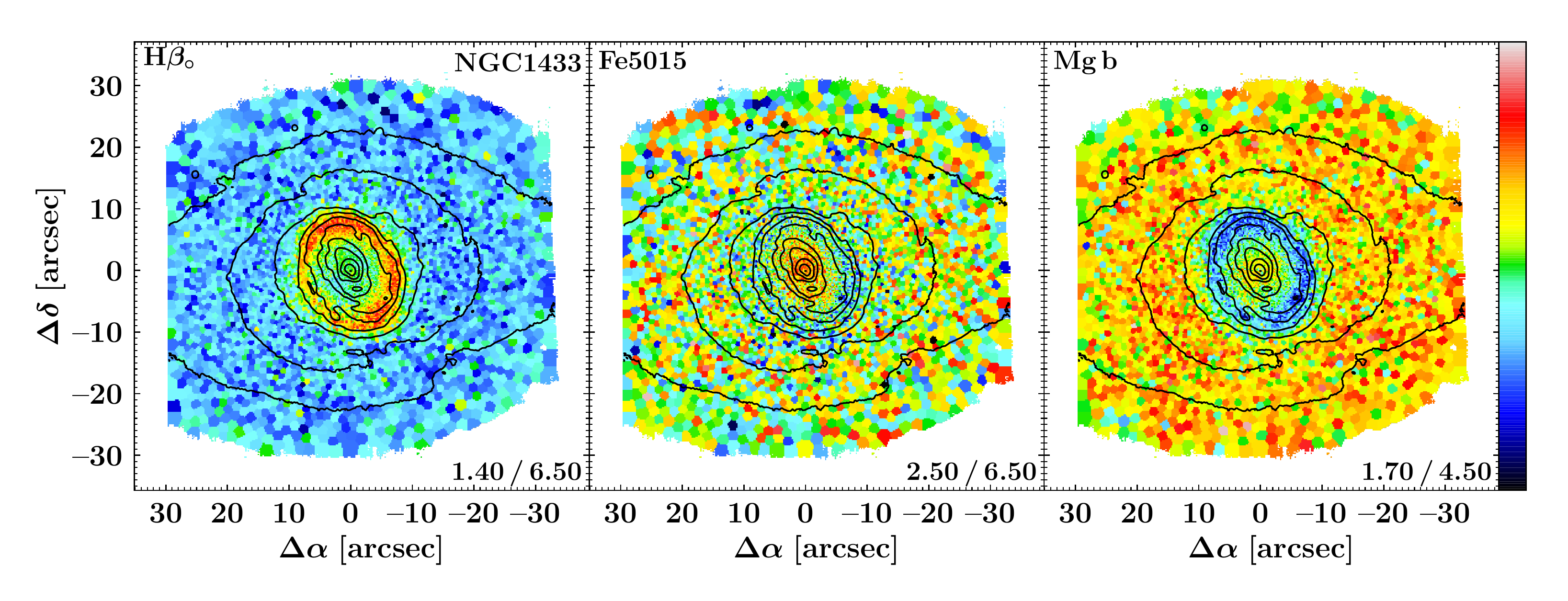}
    \caption{%
        Maps of the absorption-line strength indices H$\beta_{\circ}$ (left-hand panel), Fe5015 (central panel) and Mgb
        (right-hand panel) of the galaxy NGC\,1433.  The limits of the colourbar are stated in the lower right of each
        panel.  Isophotes are based on reconstructed intensities from the MUSE cube and displayed in steps of 0.5
        magnitudes.  North is up, east is to the left.  
    }
    \label{fig:lsIndices}
\end{figure*}

In Fig.~\ref{fig:lsIndices} we present maps of the absorption-line strength indices H$\beta_{\circ}$, Fe5015 and Mgb of
NGC\,1433.  Elevated values of the H$\beta_{\circ}$ index are only found in the nuclear ring, while its values are
constantly low at larger radii. Within the inner disc, the H$\beta_{\circ}$ index is slightly elevated as compared to
regions outside the inner disc. 
Similarly, the Fe5015 line-strength index is constant over a large part of the FoV. However, the nuclear ring becomes
evident through slightly lower index values. In the region inside the nuclear ring, the Fe5015 map shows elevated values
which further increase towards the centre. 
The Mgb line-strength index behaves similarly, with constantly high values at large radii, low values in the nuclear
ring, and elevated values that peak in the centre of the galaxy. Note that typical, median errors on the derived indices
of H$\beta_{\circ}$, Fe5015, and Mgb are approximately 0.32, 0.54 and 0.26, respectively. 

\subsection{Stellar Population Properties}
\label{subsec:populationProperties}

\subsubsection{Single Stellar Population Properties}
\begin{figure*}
    \includegraphics[width=\textwidth]{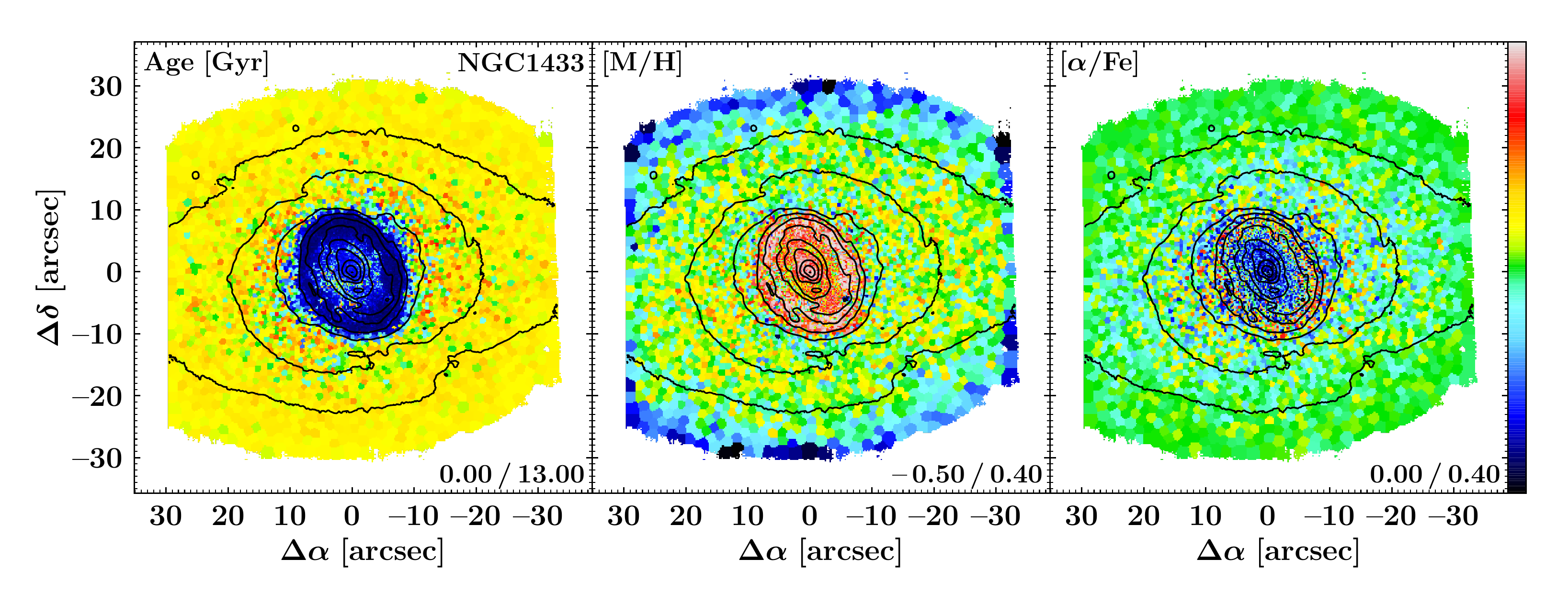}
    \caption{%
        Maps of SSP-equivalent population properties of the galaxy NGC\,1433, as derived from absorption-line strength
        measurements. Displayed are age (left-hand panel), [M/H] (centre panel) and [$\alpha$/Fe] (right-hand panel).
        The limits of the colourbar are stated in the lower right of each panel.  Isophotes are based on reconstructed
        intensities from the MUSE cube and displayed in steps of 0.5 magnitudes.  North is up, east is to the left.  
    }
    \label{fig:sspProperties}
\end{figure*}

In Fig.~\ref{fig:sspProperties} we display the SSP-equivalent population properties recovered by performing an
absorption-line strength measurement of the H$\beta_{\circ}$, Fe5015, Mgb, Fe5270, and Fe5335 indices.  We report
constant ages of $\sim$ 8~Gyr for a large part of the galaxy with the exception of the inner disc. Noteably, there we
infer younger stellar populations, reaching down to ages below 1~Gyr.  Interestingly, a region of slightly elevated ages
is found in the innermost part of the galaxy at radii smaller than $\sim$ 6~arcsec, resulting in a prominent ring-like
structure of young stellar populations at radii from $\sim$ 6 to 10~arcsec.  The spatial distribution of [M/H] shows a
similar behaviour, but with elevated values in the nuclear ring and the very centre of the galaxy. The [$\alpha$/Fe] is
constant at 0.2~dex at large radii from the centre. The nuclear ring is evident as region of higher alpha-enhancement
that is encompassed by a region of slightly less $\alpha$-enhanced stellar populations. The region inside the nuclear
ring shows significantly less $\alpha$-element enhanced stellar populations with the lowest values of $\sim$ 0.0~dex
found at the very centre. Note that typical errors on the ages, metallicities and $\alpha$-enhancements of these
recovered SSP properties are 2~Gyr, 0.15~dex, and 0.08~dex, respectively.

We highlight that the stellar population properties derived for the inner disc are clearly reminiscent of young (below
1~Gyr) stellar populations. For those young stellar populations the degeneracies between age, metallicity, and alpha
sensitive absorption-line features become severe and the conversion of line strength indices to stellar population
properties is therefore highly unreliable. Due to the recent inflow of gas towards the nuclear ring and ongoing
star-formation activity, the assumption that the stellar populations in the nuclear ring can be represented by single
age, single metallicity and single alpha-enhancement values is most likely not valid.  Nevertheless, these measurements
provide evidence that the stellar populations in the inner disc are younger and more metal rich as compared to the
remaining parts of the galaxy, while proving elevated values of alpha abundances in the nuclear ring is beyond the scope
of this study. 

\subsubsection{Non-parametric Star Formation Histories}
\label{subsubsec:sfh}
\begin{figure*}
    \includegraphics[width=\textwidth]{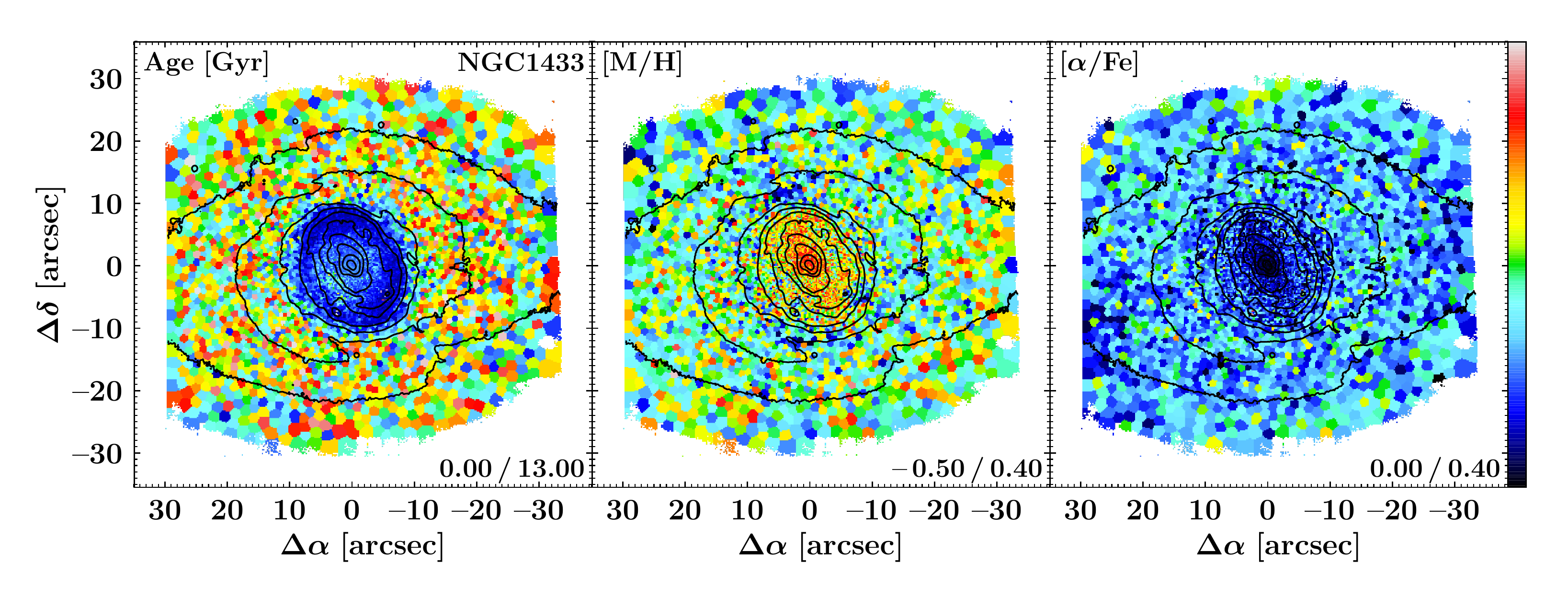}
    \caption{%
        Maps of the light-weighted stellar population content of the galaxy NGC\,1433, as derived from full-spectral
        fitting. Displayed are age (left-hand panel), [M/H] (centre panel) and [$\alpha$/Fe] (right-hand panel). The
        limits of the colourbar are stated in the lower right of each panel.  Isophotes are based on reconstructed
        intensities from the MUSE cube and displayed in steps of 0.5 magnitudes.  North is up, east is to the left.  
    }
    \label{fig:sfhProperties}
\end{figure*}

Spatial distributions of the light-weighted stellar population content, as inferred from full-spectral fitting with
\texttt{pPXF}, are presented in Fig.~\ref{fig:sfhProperties}. These results are in good agreement with the stellar
population properties derived from line strength indices above.  However, we note the difference in the $\alpha$-element
enhancement values outside of the nuclear region where the inferred SSP [$\alpha$/Fe] is ${\sim} 0.2$, instead of those
acquired through the \texttt{pPXF} full-spectral fitting $\alpha$-element enhancement of ${\sim} 0.1$.  This difference
is only slightly larger than the typical errors on the alpha-enhancement of the SSP properties (0.08~dex).  More
prominently, the ring of elevated [$\alpha$/Fe] values in the SSP maps is not found in the maps derived with
\texttt{pPXF}.  More importantly, both approaches independently converge to their lowest values of the alpha-enhancement
in the innermost regions. 

Outside the inner disc, age and [M/H] agree well, with less prominent scatter in the maps of the SSP properties.  Note
that maps of the SSP properties show slightly younger and more metal rich population in the inner disc in comparison to
the population content derived via full-spectral fitting.  Previous studies \citep[see e.g.][]{serra2007,trager2009}
have found that SSP population properties tend to show younger ages and higher metallicity as compared to those derived
from light-weighted full-spectral fitting.  In addition, SSP properties with ages below 1\,Gyr cannot be measured
reliably and the assumption of observing a \emph{single} stellar population, in particular in the inner disc, is
unrealistic. Therefore, stellar population properties in the inner disc measured with full-spectral fitting seem more
reliable and the ring of elevated [$\alpha$/Fe] values found in the SSP properties might be an unphysical result. A
comparison with mass-weighted results is beyond the scope of this study. 

\section{Summary and Conclusions}
\label{sec:conclusion}
In this study, we introduce a convenient, all-in-one framework for the scientific analysis of fully reduced,
(integral-field) spectroscopic data.  The pipeline presented in this work incorporates all necessary steps (i.e. read-in
and preparation of the science data, modular data analysis, inspection of the data products, etc.) to produce worthy
publication-quality figures.  In its default setup, the software extracts in each observed spatial position the
underlying stellar LOSVD, complemented with stellar population properties, extracted through either full-spectral or
line-strength index fitting. In conjunction, the pipeline also accounts for any ionized-gas emission lines and their
corresponding properties.  To this end, we introduce a novel Python implementation of \texttt{GandALF}. 

The \texttt{GIST} pipeline is neither specific to any IFS instrument nor a particular analysis technique and provides an
easy framework with the means of modification and development, owing to its modular code architecture. An elaborate
parallelisation is implemented and tested on machines from laptop to cluster scale.  We stress on its unique
capabilities to batch analyse data in a fully automated manner drawing on any previously produced outputs with the
possibility to interchange to a more interactive hands-on workflow facilitated by a dedicated visualisation routine with
an advanced graphical user interface.  This visualisation routine allows easy access to all measurements, spectra, fits,
and residuals, as well as star formation histories and the weight distribution of the models in fully-interactive plots. 

The presented analysis framework has successfully been applied to both low and high-redshift data from MUSE, PPAK
(CALIFA), and SINFONI, as well as to simulated data for HARMONI@ELT and WEAVE and is already being used by the TIMER,
Fornax3D, and PHANGS surveys. This further highlights the comprehensive features of the \texttt{GIST} pipeline and the
need of the scientific community for such a software package. 

We point out that this code is publicly available through a dedicated webpage\footref{webpage} and subject to ongoing
development.  Additional analysis modules, for instance the inclusion of \texttt{STECKMAP} and \texttt{pyPARADISE}, will
be included in a future release. The webpage provides a thorough documentation of the code, instructions on how to
download, install and use it, as well as a short tutorial. Because the pipeline is applicable to a large variety of
scientific objectives, we encourage the community to exploit the highly modular code architecture by adapting the
pipeline to their specific scientific needs or adding further modules.  We would also like to remind the user to
properly reference any software and spectral template libraries that have been utilised as the base of this software
package. 

We illustrate the capabilities of the pipeline by applying it to observations of NGC\,1433, obtained with the MUSE TIMER
survey.  We perform measurements of the stellar and gaseous kinematics, and infer the properties of the underlying
stellar population content.
We find evidence of a rapidly rotating inner disc, with a radial extent of approximately 10~arcsec, characterized by low
velocity dispersion and an anti-correlation between radial velocity and the h$_3$ moment. Interestingly, a correlation
between radial velocity and h$_3$ moment is also found in the innermost region of the galaxy, at radii smaller than
4~arcsec. A large amount of H$\alpha$ emission is detected, indicating gas rotation along the major axis of the galaxy
with significant departures from regular rotation.  Spatial distributions of the stellar population properties provide
evidence for a young and metal rich inner disc with low values of $\alpha$-enhancement.
These findings are all well consistent with an inner disc built by bar-driven, secular evolution processes. Forthcoming
papers on stellar kinematics and population properties of the MUSE TIMER galaxies will investigate the related processes
in detail and also connect them with the physics of gas and star-formation in the inner disc.


\begin{acknowledgements}
We thank the referee for a prompt and constructive report.  The authors thank Harald Kuntschner and Michele Cappellari
for the permission to distribute their codes together with this software package. We further thank Alexandre Vazdekis
for permission to include the MILES library. FP acknowledges Fundación La Caixa for the financial support
received in the form of a Ph.D. and post-doc contract.  FP, J. F-B and AdLC acknowledge support from grant
AYA2016-77237-C3-1-P from the Spanish Ministry of Economy and Competitiveness (MINECO).  Based on observations collected
at the European Organisation for Astronomical Research in the Southern Hemisphere under ESO programme 097.B-0640(A).
This research has made use of the SIMBAD database, operated at CDS, Strasbourg, France \citep{simbadDatabase}, NASA's
Astrophysics Data System (ADS) and the NASA Extragalactic Database (NED).  The pipeline makes use of
Astropy,\footnote{\url{http://www.astropy.org}} a community-developed core Python package for Astronomy
\citep{astropy2013, astropy2018}, as well as NumPy \citep{numpy}, SciPy \citep{scipy} and Matplotlib \citep{hunter2007}. 
\end{acknowledgements}


\bibliographystyle{aa}
\bibliography{draft}


\end{document}